\documentclass[pra,twocolumn,showpacs]{revtex4-1}

\usepackage{amsmath}
\usepackage{latexsym}
\usepackage{amssymb}
\usepackage{graphicx}
\usepackage[colorlinks=true, citecolor=blue, urlcolor=blue]{hyperref}
\usepackage{float}
\usepackage{amsfonts}
\usepackage{textcomp}
\usepackage{mathpazo}
\usepackage{comment}

\usepackage{bbm}

\usepackage{xcolor}
\definecolor{myurlcolor}{rgb}{0,0,0.6}
\definecolor{mycitecolor}{rgb}{0,0.6,0}
\definecolor{myrefcolor}{rgb}{0.6,0,0}
\usepackage{hyperref}
\hypersetup{colorlinks,
linkcolor=myrefcolor,
citecolor=mycitecolor,
urlcolor=myurlcolor}

\usepackage[draft]{fixme}
\usepackage{amsmath,bbm}%,mathrsfs}
\usepackage{graphicx}
\usepackage{amsfonts}
\usepackage{amssymb}
\usepackage{amsmath, amssymb, amsthm,verbatim,graphicx,bbm}
\usepackage{mathrsfs}
\usepackage{color,xcolor,longtable}
\newcommand{\beq}[0]{\begin{equation}}
\newcommand{\eeq}[0]{\end{equation}}

\def\be{\begin{equation}}
\def\ee{\end{equation}}
\def\ben{\begin{eqnarray}}
\def\een{\end{eqnarray}}
\def\eea{\end{array}}
\def\bea{\begin{array}}

\newcommand{\Tr}[1]{\mathrm{Tr}#1}
\newcommand{\bei}{\begin{itemize}}
\newcommand{\eei}{\end{itemize}}
\newcommand{\ket}[1]{|#1\rangle}
\newcommand{\bra}[1]{\langle#1|}

\newcommand{\proj}[1]{\ket{#1}\!\bra{#1}}

\newcommand{\N}{\mathbb{N}}
\newcommand{\Q}{\mathbb{Q}}
\newcommand{\R}{\mathbb{R}}
\newcommand{\I}{\mathbbm{1}}
\newcommand{\w}{\omega}
\newcommand{\p}{\Vec{p}}
\newcommand{\C}{\mathbbm{C}}

\renewcommand{\emph}[1]{\textbf{#1}}

\makeatletter
\newtheorem*{rep@theorem}{\rep@title}
\newcommand{\newreptheorem}[2]{%
\newenvironment{rep#1}[1]{%
 \def\rep@title{#2 \ref{##1}}%
 \begin{rep@theorem}}%
 {\end{rep@theorem}}}
\makeatother

\theoremstyle{plain}
\newtheorem*{thm*}{Theorem}
\newreptheorem{thm}{Theorem}
\newtheorem{lem}{Lemma}
\newtheorem{fakt}{Fact}
\newtheorem*{ufakt}{Fact}
\newtheorem{obs}{Observation}[lem]

\newtheorem*{udefn}{Definition}
\theoremstyle{definition}

\theoremstyle{remark}

\usepackage[T1]{fontenc}

\begin{document}
\title{Self-testing of multipartite GHZ states of arbitrary local dimension\\ with arbitrary number of measurements per party}
\author{Shubhayan Sarkar}
\affiliation{Center for Theoretical Physics, Polish Academy of Sciences, Aleja Lotnikow 32/46, 02-668 Warsaw, Poland }
\author{Remigiusz Augusiak}
\affiliation{Center for Theoretical Physics, Polish Academy of Sciences, Aleja Lotnikow 32/46, 02-668 Warsaw, Poland }

\begin{abstract}
Device independent certification schemes have gained a lot of interest lately, not only for their applications in quantum information tasks but also their implications towards foundations of quantum theory. The strongest form of device independent certification, known as self-testing, often requires for a Bell inequality to be maximally violated by specific quantum states and measurements. In this work, using the techniques developed recently in [S. Sarkar \textit{et al.}, \href{https://www.nature.com/articles/s41534-021-00490-3}{npj Quantum Inf. \textbf{7}, 151 (2021)}], we provide the first self-testing scheme for the multipartite Greenberger–Horne–Zeilinger (GHZ) states of arbitrary local dimension that does not rely on self-testing results for qubit states and that exploits the minimal number of two measurements per party. This makes our results interesting as far as practical implementation of device-independent certification methods is concerned.  Our self-testing statement relies on maximal violation of a Bell inequality proposed recently in [R. Augusiak \textit{et al.}, \href{https://iopscience.iop.org/article/10.1088/1367-2630/ab4d9f}{New J. Phys. \textbf{21}, 113001 (2019)}]. 
%which is violated by arbitrary number of measurements per site, making our self-testing scheme much %more applicable in real-life experiments.  
\end{abstract}

\maketitle

\section{Introduction}
Quantum theory has presented us with numerous counter-intuitive predictions, most of which have been verified by experiments till date. Many of them have been shown to have no classical analogue. One such prediction of quantum theory is the existence of certain correlations that arise by performing local measurements on a joint entangled quantum state, which violate assumptions which any classical theory must abide. Such correlations are termed Bell non-local or simply non-local and are detected by violating Bell inequalities \cite{Bell,Bell66}. Interestingly, non-local correlations constitute a powerful resource for numerous applications, in particular within the device-independent framework in which 
one does not need to make any assumptions on the devices used to perform a given task
except that they follow the rules of quantum theory. A prominent example of such applications
is the device-independent quantum cryptography \cite{di2} (see also Ref. \cite{bellreview}).

Another such application that has gained a lot of interest within the quantum community is device-independent certification. It is actually an umbrella term encompassing a few tasks whose general aim is to make nontrivial statements about the underlying quantum system based only on the observed non-local correlations. This last fact makes device-independent certification schemes interesting from the practical point of view as they require much less information about the underlying system to deduce its relevant properties. For instance, Bell non-locality has been shown to enable DI certification of quantum system's dimension \cite{dimension}, that a given state is entangled \cite{DIEW} or even the amount of entanglement present in it \cite{Moroder}. It allows to certify that the outcomes of quantum measurements are truly random \cite{di4,di3}.

The most fascinating and at the same time most complete form of device-independent certification is
self-testing. Introduced in \cite{di1}, it aims to harness the observed quantum correlations to provide almost full characterization of the underlying joint quantum state as well as the measurements performed on it; here, almost refers to the fact that, being based on the obtained statistical data, such certification can only be made up to certain undetectable degrees of freedom such as invariance under the action of local unitary transformation or adding an extra system that gives no contribution to the observed non-locality. Within recent years there has been a substantial effort to propose self-testing schemes for various quantum states and/or measurements
such as for instance those in Refs. \cite{Scarani,Yang,Bamps,All,chainedBell,Coladangelo,Laura,Jed,Sarkar,Jed2,Mckay,graph}. However, most of them have been designed for bipartite entangled systems, whereas the multipartite scenario remains highly unexplored, in particular when quantum systems of arbitrary local dimension are concerned. The existing multipartite methods are devised for $N$-partite graph states \cite{Jed2, Mckay, graph} or Dicke states \cite{Wstate,Remik2}, which are nevertheless locally qubits. To the best of our knowledge, the only known multipartite self-testing scheme for multipartite states of local dimensions higher than two concerns the so-called Schmidt states including the well-known $N$-qudit the Greenberger-Horne-Zeilinger (GHZ) state \cite{Remik2},
\begin{eqnarray}\label{GHZstatem}
\ket{\text{GHZ}_{N,d}}=\frac{1}{\sqrt{d}}\sum_{i=0}^{d-1}\ket{i}^{\otimes N}
\end{eqnarray}
with $N$ and $d$ being arbitrary integers such that $N,d\geq 2$.
However, this scheme, being an adaptation of the results of Ref. \cite{Coladangelo} to the multiparty scenario, relies on application of many self-testing schemes for the two-qubit states, and requires the parties to perform three or four measurements in order to certify the state. It is thus a vital problem in the domain of DI certification whether it is possible to design schemes that require 
less effort to be practically implemented.

The main aim of our work is to provide a self-testing strategy for the $N$-partite GHZ
states of local dimension $d$, which is based on maximal violation of a Bell inequality involving an arbitrary number of truly $d$-outcome measurements. Moreover, in the simplest case, our scheme requires measuring only two observables at each site, which is in fact the minimal number of measurements necessary to observe quantum non-locality and thus to make nontrivial self-testing statements. On the other hand, we generalize some previous results in a few ways: (i) 
first, our results generalize the recent self-testing statement 
for the two-qudit maximally entangled states \cite{Sarkar} to an arbitrary number of parties
as well as an arbitrary number of measurements; (ii) it also generalizes the results
of Ref. \cite{chainedBell} derived for the chained Bell inequalities to an arbitrary number of parties and an arbitrary local dimension.

\section{Preliminaries} 
Before getting to results, let us first describe the scenario and 
introduce the relevant notions.

\subsection{Multipartite Bell scenario}

We consider here the multipartite Bell scenario comprising $N$ 
spatially separated parties, denoted $A_i$ $(i=1,\ldots,N)$,
and one preparation device distributing among them an $N-$partite state $\rho_N$ that acts on
$\mathcal{H}_1\otimes\ldots\otimes\mathcal{H}_N$ with each $\mathcal{H}_i$ being a finite-dimensional Hilbert space representing the physical system of party $A_i$. 
On their shares of the state the observers perform measurements and register the obtained outcomes. 
We consider here a general scenario in which each party $A_i$ can freely choose to perform one of $m$ measurements, each having $d$ outcomes, where both $m$ and $d$ are arbitrary. The measurements are denoted $M_{i,x_i}$ with $x_i=1,\ldots,m$ labelling the measurement choices of party $A_i$, whereas the outcomes are denoted $a_i$ with $a_i=0,\ldots,d-1$. 

The correlations observed by the parties are encoded into a vector of joint probability distributions, %
\begin{equation}\label{probdist}
    \p = \{p(a_1,\ldots,a_N|x_1,\ldots,x_N)\}\in \R^{(md)^N}.
\end{equation}
where $p(a_1,\ldots,a_N|x_1,\ldots,x_N)\equiv p(\boldsymbol{a}|\boldsymbol{x})$ denotes the joint probability of obtaining $a_i$ by the party $A_i$ after performing the measurement $x_i$ and is given by the well-known Born's formula
%
%As mentioned above, the parties are spatially separated which points to the fact the there is no %communication between them. Mathematically, this can be understood as some set of constraints known %as no-signalling conditions which the above joint probability distributions need to obey, 
%
%\begin{eqnarray}
%\sum_{a_i}p(a_1,a_2,\ldots a_i,\ldots,a_N|x_1,x_2,\ldots %x_i,\ldots,x_N)\nonumber\\=\sum_{a_i}p(a_1,a_2,\ldots a_i,\ldots,a_N|x_1,x_2,\ldots x_i',\ldots,x_N)
%\end{eqnarray}
%
%for all $x_i, x_i'$. The set of joint probability distribution obeying the above constraint form a %convex set known as set of no-signalling correlations denoted by $\mathcal{N}_{m,d,N}$. 
%%For details refer to \cite{bellreview}. 
%
%Now, if the underlying descriptions which give rise to the above joint probability distributions can %be attributed to quantum theory, then we have a well known expression for calculating the joint %probability given by
%
\begin{eqnarray}
&&p(a_1,\ldots,a_N|x_1,\ldots,x_N)\nonumber\\
&&\hspace{1cm}=\Tr\left[\rho_{N}\left(M_{1,x_1}^{a_1}\otimes\ldots\otimes M_{N,x_N}^{a_N}\right)\right],
\end{eqnarray}
where $M_{i,x_i}^{a_i}$ are the measurement operators corresponding to the outcome $a_i$ of the measurement $x_i$; recall that these are positive semi-definite and satisfy $\sum_{a_i}M_{i,x_i}^{a_i}=\I$ for all $x_i$ and $i$. The set of joint probability distributions achievable using quantum states and quantum measurements is usually referred to as the set of quantum correlations or simply the quantum set; we denoted it by $\mathcal{Q}_{N,m,d}$.
%
%which is also convex and a proper subset of $\mathcal{N}_{m,d,N}$. 

Let us consider a certain subset of the set of quantum correlations $\mathcal{Q}_{m,d,N}$ which can be represented using local-realistic descriptions of the underlying system, commonly referred to as the set of classical or local correlations, denoted $\mathcal{L}_{m,d,N}$. Precisely, the latter contains those correlations that admit the following representation
\begin{equation}
p(a_1,\ldots,a_N|x_1,\ldots,x_N)=\sum_{\lambda}\mu(\lambda) \prod_{i=1}^Np(a_i|x_i,\lambda),
\end{equation}
where $\lambda$ is a random variable distributed according to a distribution $\mu(\lambda)$ and $p(a_i|x_i,\lambda)\in\{0,1\}$ for every $x_i$, $a_i$ and $i$. Similarly to $\mathcal{Q}_{N,m,d}$, the set $\mathcal{L}_{N,m,d}$ is convex; in fact, it is a polytope for any choice of $N$, $m$ and $d$. 

%%%%%%%%%%%%%%%%%%%%%%%%%
\subsection{Bell inequalities.} 

As it was first observed by Bell in 1964 \cite{Bell}, in the scenario with two parties, each performing two $2-$outcome measurements the local set $\mathcal{L}_{2,2,2}$ is a proper subset of $\mathcal{Q}_{2,2,2}$. %This was the first observation of non-local correlations or correlations which cannot be simulated %using classical strategies. 
To this end, he considered certain inequalities that are linear in $p(\boldsymbol{a}|\boldsymbol{x})$ that constrain the set of local correlations. These are typically termed Bell inequalities
and their general form reads
%
%Taking into account the fact that the local set is a polytope a natural way to reveal that a given %quantum behaviour $\vec{p}$ does not belong to the local set is to use Bell inequalities
%whose general form is given by
%
\begin{eqnarray}\label{BI}
\mathcal{I}: = \Vec{t}\cdot\p &\leq& \beta_L, 
\end{eqnarray}
where 
\begin{equation}
\vec{t}=\{t_{a_1,\ldots,a_N,x_1,\ldots,x_N}\}    
\end{equation}
is a vector consisting of real numbers, $\vec{t}\in\R^{(md)^N}$. 
The number appearing on the right-hand side of (\ref{BI}) is the maximal value of 
the Bell expression $\mathcal{I}$ over all classical strategies, $\beta_L=\max_{\p\in\mathcal{L}_{m,d,N}} \mathcal{I}$, 
and is typically referred to as the classical or local bound. Analogously, by $\beta_Q=\sup_{\p\in\mathcal{Q}_{m,d,N}}\mathcal{I}$ one denotes the maximal value of $\mathcal{I}$ achievable by quantum correlations and refers to it as the quantum or the Tsirelson bound; the quantum set is in general not closed \cite{Slofstra,Beigi}, which explains the supremum in the definition of $\beta_Q$. 

Importantly, violation of (\ref{BI}) by some $\vec{p}$ implies that the latter is non-local. Moreover, if $\vec{p}$ violates a Bell inequality maximally or, in other words, achieves the maximal quantum value $\beta_Q$, then it necessarily lies at the boundary of the corresponding quantum set $\mathcal{Q}_{N,m,d}$.

In what follows it will be more convenient for us  
to express Bell inequalities in terms of correlators instead of probabilities.
Due to the fact that here we deal with quantum measurements with an arbitrary number of outcomes, 
we will use generalized expectation values, which are in general complex numbers
defined as the multidimensional Fourier transform of $p(\boldsymbol{a}|\boldsymbol{x})$ (see, e.g., Refs. \cite{SATWAP,ACTA}):
%
%describe correlations observed in  Bell experiments is to use expectation values of joint observables %of different parties denoted by $\langle \mathcal{A}_{1,x_1}\mathcal{A}_{2,x_2}\ldots %\mathcal{A}_{N,x_N} \rangle$. Both representations of Bell inequalities are inter-convertible using %the fourier transform given as
%
\begin{equation}\label{ExpValues}
    \left\langle A_{1,x_1}^{k_1}\ldots A_{N,x_N}^{k_N} \right\rangle = \sum_{a_1,\ldots,a_N=0}^{d-1}    \omega^{\boldsymbol{a}\cdot\boldsymbol{k}} p(\boldsymbol{a}|\boldsymbol{x}),
\end{equation}
where $\w$ is the $d$-th root of unity $\omega=\exp(2\pi\mathbbm{i}/d)$ and $\boldsymbol{k}$ is an $N$-component vector composed of $k_i=0,\ldots,d-1$ for all $i$, and, finally, $\boldsymbol{a}\cdot\boldsymbol{k}=a_1k_1+\ldots a_Nk_N$ stands for the standard scalar product of two real vectors. 

Crucially, if the measurements performed by the observers are projective, the expectation values (\ref{ExpValues}) can be represented as 
\begin{equation}
\left\langle A_{1,x_1}^{k_1}\ldots A_{N,x_N}^{k_N} \right\rangle =\Tr\left[\left(A_{1,x_1}^{k_1}\otimes\ldots\otimes A_{N,x_N}^{k_N}\right)\rho_{N}\right],
\end{equation}
where now $A_{i,x_i}$ are unitary operators defined as one-dimensional Fourier transforms of the measurement operators $M_{x_i}^{a}$,
\begin{equation}
    A_{i,x_i}^{k_i} = \sum^{d-1}_{a_i=0} \w^{a_ik_i} M^{a_i}_{i,x_i}, 
\end{equation}
for $k_i=0,1,\ldots,d-1$ and $i=1,\ldots,N$. Due to the fact that for projective measurements
the operators $M^{a_i}_{i,x_i}$ are pairwise orthogonal
%
%, that is, $M^{a_i}_{i,x_i}M_{i,x_i}^{a_i'}=\delta_{a_i,a_i'}M^{a_i}_{i,x_i}$ 
%
for any $x_i$ and $i$, it is not difficult to realize that all $A_{i,x_i}^{k_i}$ are unitary operators with eigenvalues $\omega^i$ for $i=0,\ldots,d-1$. Moreover, $A_{i,x_i}^{k_i}$ is 
simply the $k_i$th power of $A_{i,x_i}$; in what follows we refer to $A_{i,x_i}$ as quantum observables.

In fact, as discussed below, since we are concerned with self-testing we can safely restrict our attention to projective measurements; on the same basis we can also assume that the shared state is pure.

%It should be noticed here that since we work with self-testing, the local Hilbert spaces %$\mathcal{H}_i$ are of unknown dimension and therefore we can safely assume the measurements 
%to be projective. Analogously, we can assume that the distributed state is pure, i.e., %$\rho_N=\proj{\psi_N}$ for some $\ket{\psi_N}\in\mathcal{H}_1\otimes\ldots\otimes\mathcal{H}_N$. %Using the above descriptions, we are now ready to introduce the idea of self-testing.
%
%it is fully justified to restrict our attention projective measurements Note that in the above %scenario we do not make assumptions on the dimensions of the local Hilbert spaces and therefore, %without any loss of generality, we can assume that the local measurements are projective as well as that the joint state is pure, i.e., $\rho_N=\proj{\psi_N}$. Using the above descriptions, we are now ready to introduce the idea of self-testing.

%%%%%%%%%%%%%%%%%%
\subsection{Self-testing}

%We are now ready to present our main result: the self-testing statement for the two-qudit maximally entangled state and certain $d$-outcome measurements usually referred to as Collins-Gisin-Linden-Massar-Popescu (CGLMP) measurements \cite{Zukowski,CGLMP,BKP} (see Appendix \ref{sec:cglmp} for their explicit form).

Let us finally define the task of self-testing. To this end, we consider again the Bell experiment described above, assuming this time that the functioning of all the devices involved is unknown. 
That is, the parties have no knowledge about the shared state $\rho_N$ as well as the corresponding Hilbert space, and they do not know the measurements their devices perform; in fact these can be treated as black boxes which when supplied with an input $x_i=1,\ldots,m$ return one of possible outputs $a_i=0,\ldots,d-1$. Yet, the measuring devices are assumed to behave according to the rules of quantum theory. Now, due to the fact that the local Hilbert spaces are uncharacterized 
we can make here the standard assumption that the shared state is pure, that is, $\rho_N=\proj{\psi_N}$ for some $\ket{\psi_N}\in\mathcal{H}_N$ and that the measurements are projective.

%
%when  distributed to them assume this time that the preparation device prepares a state $\rho_N$ %that remains unknown  and assume that the shared state $\rho_N$ as well as the measurements %performed by the parties are unknown to them.
%This time we additionally assume that the state shared by 
%

Now, based on the observed correlations represented by $\vec{p}$ or, 
equivalently, by the expectation values (\ref{ExpValues}), the parties aim at 
making nontrivial statements about the state $\rho_N$ and/or the 
measurements $A_{i,x_i}$ performed on it. This general task is usually referred to as
device-independent certification. Its strongest form is self-testing in which the parties 
use the observed correlations to certify that the shared state $\ket{\psi_N}$ 
as well as the measurements performed on it are equivalent, 
up to some well-understood equivalences, to some known
state $\ket{\psi_N'}\in(\mathbbm{C}^d)^{\otimes N}$ 
and known observables $A_{i,x_i}'$ acting on $\mathbbm{C}^d$. 
To be more precise, let us formulate the following definition.
\begin{udefn}
We say that the observed correlations 
self-test the state $\ket{\psi_N'}$ and the observables $A_{i,x_i}'$ if one 
can deduce that: each local Hilbert space decomposes as $\mathcal{H}_i=\mathbbm{C}^d\otimes\mathcal{H}_{i}''$ for some finite-dimensional 
$\mathcal{H}_{i}'$, and (ii) there exist local unitary operations 
$U_i:\mathcal{H}_i\to\mathcal{H}_i$ such that 
\begin{equation}
    U_1\otimes\ldots\otimes U_N\ket{\psi_N}=\ket{\psi_N'}\otimes \ket{\mathrm{aux}_N}
\end{equation}
for some auxiliary state $\ket{\mathrm{aux}_N}\in\mathcal{H}'_1\otimes\ldots\otimes\mathcal{H}'_{N}$
and
\begin{equation}
    U_i\,A_{i,x_i}\, U_i^{\dagger}=A_{i,x_i}'\otimes\mathbbm{1}_i,
\end{equation}
where $\mathbbm{1}_i$ is an identity acting on $\mathcal{H}_{i}'$.
\end{udefn}

Notice that in our case the reference state $\ket{\psi_N'}$ is the GHZ state
(\ref{GHZstatem}), whereas the reference observables $A_{i,x_i}'$ are provided explicitly below 
in Eqs. (\ref{measurementsm}), (\ref{measurements2m}) and (\ref{measurements3m}).

%almost fully characterise the preparation $\ket{\psi_N}$ 
%as well as the performed measurements $A_{i,x_i}$, where 'almost' refers to the fact that 
%the correlations $\vec{p}=\{p(\boldsymbol{a}|\boldsymbol{x})\}$ are invariant under 
%certain degrees of freedom that cannot be detected by the parties. 
%
%can reveal the underlying states and measurements up to the degrees of freedom which do not change %the joint probability distribution \eqref{probdist}. 
%
%To be more precise, let us formulate the following definition 

%let us name these degrees of freedom. First, the set of local unitaries $U_{\mathcal{A}_i}$ that act %on $\mathcal{H}_{\mathcal{A}_i}$ for all $i$, that is, the state $U_{\mathcal{A}_1}\otimes %U_{\mathcal{A}_2}\ldots U_{\mathcal{A}_N}|\psi_{N}\rangle$ together with the measurements %$\{U_{\mathcal{A}_i}\mathcal{A}_{i,x_i}U^\dagger_{\mathcal{A}_i}\}$ for all $x_i$. Second, due to no %assumption on dimension of the state, there might exist some auxiliary system on which the %measurements act trivially and is uncorrelated with the state which generates the statistics.

A necessary condition to derive a self-testing statement based on the observed correlations $\vec{p}$ is that they violate some Bell inequality maximally. Hence, the first task is to identify a Bell inequality that is maximally violated by the multipartite GHZ state of arbitrary dimension with arbitrary number of measurements per party. Quite recently, a Bell inequality meeting this requirement was derived in \cite{ACTA}. In the correlator picture it can be stated in the following form
\begin{eqnarray}\label{eqASTA}
\left\langle\mathcal{\hat{I}}_{N,m,d}\right\rangle\leq \beta_L,
\end{eqnarray}
where $\mathcal{\hat{I}}_{N,m,d}$ is the Bell operator given by
\begin{eqnarray}\label{eqASTAopm}
\mathcal{\hat{I}}_{N,m,d}:&=&\sum_{\alpha_1,\ldots,\alpha_{N-1}=1}^{m}\sum_{k=1}^{d-1}\left(a_kA_{1,\alpha_1}^k\otimes\bigotimes_{i=2}^N A^{(-1)^{i-1}k}_{i,\alpha_{i-1}+\alpha_i-1}\right.\nonumber\\\hspace{1cm}&&+\left.a_k^*A_{1,\alpha_1+1}^k\otimes\bigotimes_{i=2}^NA^{(-1)^{i-1}k}_{i,\alpha_{i-1}+\alpha_i-1}\right),
\end{eqnarray}
where the complex coefficients $a_k$ are given by
\begin{equation}
a_k=\frac{\omega^{(2k-d)/4m}}{2\cos(\pi/2m)},
\end{equation}
and we assume the convention that $A_{i,m+1}=\omega A_{i,1}$ and $\alpha_N=1$. 

The maximal quantum value of this inequality is known to be $\beta_Q=m^{N-1}(d-1)$. At the same time, the local bound $\beta_L$ has been computed numerically only for some cases in \cite{ACTA}; yet, it was shown that $\beta_L<\beta_Q$ for all finite $N$ and $d$. %(\textbf{Remik: check}). 
The maximal quantum value can be achieved by the $N$-partite GHZ state of local dimension $d$ defined in Eq. \eqref{GHZstatem} and the following measurements 
\begin{equation}\label{measurementsm}
\mathcal{O}_{1,x}=U_{x}F_d\,\Omega_dF_d^{\dagger}U_{x}^{\dagger},\qquad
\mathcal{O}_{2,x}=V_{x}F_d^{\dagger}\Omega_dF_dV_{x}^{\dagger},
\end{equation}
for the first two parties, and
\begin{equation}\label{measurements2m}
\mathcal{O}_{\mathrm{odd},x}=W_{x}F_d\,\Omega_dF_d^{\dagger}W_{x}^{\dagger}
\end{equation}
and
\begin{equation}\label{measurements3m}
\mathcal{O}_{\mathrm{ev},x}=W_{x}^{\dagger}F_d^{\dagger}\Omega_dF_d W_{x}    
\end{equation}
for all other parties $A_i$ with $i=3,\ldots,N$ numbered by odd and even numbers, respectively.
Here
\begin{equation}
F_d=\frac{1}{\sqrt{d}}\sum_{i,j=0}^{d-1}\omega^{ij}\ket{i}\!\bra{j},\qquad
\Omega_d=\mathrm{diag}[1,\omega,\ldots,\omega^{d-1}]
\end{equation}
with $\omega=\mathrm{exp}(2\pi\mathbbm{i}/d)$. Then, the unitary operations 
$U_x$, $V_x$ and $W_x$ are defined as
\begin{equation}
U_{x}=\sum_{j=0}^{d-1}\omega^{-j \alpha_{M}(x)}\proj{j},\qquad
V_x=\sum_{j=0}^{d-1}\omega^{j \beta_{M}(x)}\proj{j},
\end{equation}
and
\begin{equation}
    W_x=\sum_{j=0}^{d-1}\omega^{-j \gamma_{M}(x)}\proj{j},
\end{equation}
where 
\begin{equation}\label{measuparam}
\gamma_m(x)=\frac{1}{m}\left(x-\frac{1}{2}\right),\quad \zeta_m(x)=\frac{x}{m},\quad \theta_m(x)=\frac{x-1}{m}.
\end{equation}
It is worth noticing that the above observables are by the very definition unitary and that their eigenvalues are $\omega^{i}$ with $i=0,\ldots,d-1$, and thus they perfectly match our scenario. Moreover, for the particular case $N=m=2$, they reproduce the well-known Collins-Gisin-Linden-Massar-Popescu (CGLMP) measurements \cite{CGLMP, BKP}.

In the next section, we prove that the above mentioned state and the measurements are the only realisations up to the freedom of local unitaries and some auxiliary system which can saturate the quantum bound of the inequality \eqref{eqASTA}.

%%%%%%%%%%%%%%%%%%%%%%%%%
\section{Results}

We are now ready to present our main result, that is, the self-testing statement for
the GHZ state (\ref{GHZstatem}) of arbitrary local dimension and the corresponding measurements (\ref{measurementsm}), (\ref{measurements2m}) and (\ref{measurements3m}). 
The key ingredient in establishing this result is a sum-of-squares decomposition of the Bell operator $\hat{\mathcal{I}}_{N,m,d}$ provided in Ref. \cite{ACTA}. Indeed, for any choice of the local observables $A_{i,x_i}$ acting on $\mathcal{H}_i$ one has 
\begin{eqnarray}\label{SOS1}
\hspace{-0.5cm}\beta_Q\I-\mathcal{\hat{I}}_{N,m,d}=\frac{1}{2}\sum_{\alpha_1,\ldots\alpha_N=1}^m\sum_{k=1}^{d-1}\left(P_{\alpha_1,\ldots,\alpha_N}^{(k)}\right)^{\dagger}P_{\alpha_1,\ldots,\alpha_N}^{(k)}\nonumber\\
+\frac{m^{N-2}}{2}\sum_{\alpha=1}^{m-2}\sum_{k=1}^{d-1}\left(R_{\alpha}^{(k)}\right)^{\dagger}R_{\alpha}^{(k)},
\end{eqnarray}
where
\begin{equation}\label{P}
P_{\alpha_1,\ldots,\alpha_N}^{(k)}=\I-\left(a_k A_{1,\alpha_1}^k+a_k^*A_{1,\alpha_1+1}^k\right)\otimes\bigotimes_{i=2}^N
A^{(-1)^{i-1}k}_{i,\alpha_{i-1}+\alpha_i-1}
\end{equation}
and
\begin{eqnarray}
R_{\alpha}^{(k)}=\mu_{\alpha,k}^*\, A_{1,2}^k+\nu_{\alpha,k}^* \,A_{1,\alpha+2}^k+\tau_{\alpha,k}\,A_{1,\alpha+3}^k
\end{eqnarray}
for $\alpha=1,\ldots,m-2$ and $k=1,\ldots,d-1$, and $\mathbbm{1}$.
Then, the coefficients $\mu_{\alpha,k}, \nu_{\alpha,k}$ and $\tau_{\alpha,k}$ can be found in Appendix A.

Before stating our main result let us introduce two unitary observables with eigenvalues $\omega_i$. The first one is the $d-$dimensional generalization of $\sigma_z$-Pauli matrix given by
\begin{eqnarray}
Z_d=\sum_{i=0}^{d-1}\omega^{i}\ket{i}\!\bra{i}
\end{eqnarray}
whereas the second one is defined as
\begin{eqnarray}
\label{ZdTd1}
%Z_d&=&\sum_{i=0}^{d-1}\omega^i\proj{i}\nonumber,\\ 
T_{d,m}&=&\sum_{i=0}^{d-1}\omega^{i+\frac{1}{m}}\proj{i}-\frac{2\mathbbm{i}}{d}\sin\left(\frac{\pi}{m}\right)\nonumber\\
&&\times\sum_{i,j=0}^{d-1}(-1)^{\delta_{i,0}+\delta_{j,0}}\omega^{\frac{i+j}{2}-\frac{d-2}{2m}}|i\rangle\!\langle j|,
\end{eqnarray}
where $\delta_{i,j}$ is the Kronecker delta such that $\delta_{i,j}=1$ for $i=j$
and $\delta_{i,j}=0$ otherwise. Note that for $m=d=2$,  $T_{2,2}=-\sigma_x$, which is another Pauli matrix.

Now, we can state our main theorem.
\begin{thm*}
Assume that the Bell inequality (\ref{eqASTA}) is maximally violated by some state $\ket{\psi_{N}}\in\mathcal{H}_{1}\otimes\ldots\otimes\mathcal{H}_{N}$ and unitary observables $A_{i,\alpha_i}$ for $i\in \{1,2,\ldots,N\}$ and $\alpha_i\in \{1,2,\ldots,m\}$. Then, each local Hilbert space decomposes as $\mathcal{H}_i=\mathbbm{C}^d\otimes \mathcal{H}_{i}''$ for some finite-dimensional Hilbert spaces $\mathcal{H}_i''$, and there exist local unitary transformations
%
%\begin{equation}
 $   U_{i}:\mathcal{H}_{i} \rightarrow \mathcal{H}_{i}$
such that 
\begin{eqnarray}\label{AiBimain1}
U_{i} A_{i,\alpha_i} U^{\dagger}_{i} &=& \mathcal{O}_{i,\alpha_i} \otimes \I_{i}''
\end{eqnarray}  
where $\mathcal{O}_{i,\alpha_i}$ are the $d\times d$ observables defined in Eqs. \eqref{measurementsm}, \eqref{measurements2m} and \eqref{measurements3m}, and $\I_{i}''$ are the identity matrices acting on $\mathcal{H}_{i}''$ for all $i's$, and, finally,
%
%Moreover, the state $\ket{\psi_{N}}$ which gives the maximal violation can be stated upto the freedom %of local unitaries and the auxiliary state as,
%
\be\label{state}
U_{1}\otimes \ldots\otimes U_{N}|\psi_{N}\rangle=\ket{\mathrm{GHZ}_{N,d}} \otimes \ket{\mathrm{aux}_{N}}, \ee 
for some $\ket{\mathrm{aux}_{N}}\in \mathcal{H}_{1}''\otimes\ldots \otimes \mathcal{H}_{N}''$.
\end{thm*}
\begin{proof}The proof is very technical and long and therefore here we present its sketch, whereas its complete version is deferred to Appendix \ref{sec:selftesting}. 
It is divided into two major parts. In the first part of the proof, we begin by concentrating on the first party and 
characterise its Hilbert space as well as observables; in fact, we show that 
in $\mathcal{H}_1$ one can identify a qudit Hilbert space $\mathbbm{C}^d$ and
prove the existence of a unitary operation that brings all $A_{1,x}$ to the ideal measurements
(\ref{measurementsm}). Then, we extend the above observations to the remaining parties. In the second part of the proof, we exploit the obtained observables to show that $\ket{\psi_N}$ is unitarily equivalent to the $N$-qudit GHZ state.
%plug in the obtained measurements and show that the only state which satisfies \eqref{SOSrel2} is %$N$-partite GHZ state of local dimension $d$ \eqref{GHZstatem}, uncorrelated with some auxiliary %state on which the measurements act trivially. Due to the technicality of the proof, we address here %only the major steps of the proof with all the details provided in Appendix A.

%each local Hilbert space $\mathcal{H}_{i}$ 
%
%on which the measurements $\mathcal{A}_{i,\alpha_i}$ act for all $i, \alpha_i$ 
%

\textit{The Hilbert spaces structure and characterisation of observables.} Let us begin by showing that maximal violation of the Bell inequality (\ref{eqASTA}) allows one to identify a qudit in each local Hilbert space in the sense that $\mathcal{H}_{i}= \C^d \otimes \mathcal{H}_{i}''$ for any $i$, and, simultaneously, to obtain the form of $A_{i,2}$ and $A_{i,3}$ for any $i=2,\ldots,N$.

We concentrate on the first party $A_1$, the proof for the other $A_i$'s follow exactly the same lines. The departure point for our considerations are certain relations for the observables $A_{i,x_i}$ and the state $\ket{\psi_N}$ that are induced by the sum-of-squares decomposition 
(\ref{SOSrelam}). Precisely, this decomposition implies that any $\ket{\psi_{N}}$ and $A_{i,x_i}$ maximally violating the Bell inequality (\ref{eqASTA}) must necessarily satisfy 
\begin{equation}
    P_{\alpha_1,\ldots,\alpha_N}^{(k)}\ket{\psi_{N}}=0
\end{equation}
for any configuration of the indices $\alpha_i$, which through (\ref{P})
implies that 
\begin{eqnarray}\label{SOSrel2m}
\overline{A}_{1,\alpha_1}^{(k)}\otimes\bigotimes_{i=2}^N A^{(-1)^{i-1}k}_{i,\alpha_{i-1}+\alpha_i-1}\ket{\psi_{N}}=\ket{\psi_{N}}
\end{eqnarray}
for all $k$ and $\alpha_i$, where we have denoted 
\begin{equation}\label{comb}
    \overline{A}_{1,\alpha_1}^{(k)}=a_k A_{1,\alpha_1}^k+a_k^*A_{1,\alpha_1+1}^k.
\end{equation}
Since $A_{i,\alpha_{i}}$ are unitary for all $i$ and $\alpha_i$, we then straightforwardly conclude that the operators acting on the first party's Hilbert space must satisfy the following relations 
\begin{eqnarray}\label{SOSrelam}
\overline{A}_{1,\alpha_1}^{(k)}\overline{A}_{1,\alpha_1}^{(d-k)}=\I_{A_1}, \quad  \text{and} \quad \overline{A}_{1,\alpha_1}^{(k)}=\left[\overline{A}_{1,\alpha_1}^{(1)}\right]^k
\end{eqnarray}
for any $k=1,\ldots,d-1$ and $\alpha_1=1,\ldots,m$. By noting that $a_{d-k}=a_{k}^{*}$, one further obtains 
\begin{equation}
\overline{A}_{1,\alpha_1}^{(d-k)}=\overline{A}_{1,\alpha_1}^{(k)\dagger}
\end{equation}
for all $k,\alpha_1$, and therefore the relations (\ref{SOSrelam})
imply that the combinations of observables (\ref{comb}) are also quantum observables, i.e., 
are unitary and their spectra are from $\{1,\w,\ldots,\omega^{d-1}\}$
Additionally, we have another set of relations arising from the SOS decomposition \eqref{SOS1} given by 
\begin{equation}
    R_{\alpha}^{(k)}\ket{\psi_{N}}=0
\end{equation}
which due to the fact that the single-site reduced density matrices of the state $\ket{\psi_N}$ are full rank, are equivalent to
\begin{eqnarray}\label{SOSrelbm}
R_{\alpha}^{(k)}=0
\end{eqnarray}
for all $k=1,\ldots,d-1$ and $\alpha=1,\ldots,m-2$. The relations \eqref{SOSrelam} and \eqref{SOSrelbm} are key factors in proving our self-testing statement. 

In fact, using them we can prove the following lemma. 

%%%%%%%%%%%%%%%%%%%%%%%%%%

\begin{lem}\label{le:traceless}
Consider the unitary observables $A_{1,\alpha_1}$ for $\alpha_1= 2,3$ acting on $\mathcal{H}_1$ whose eigenvalues are $\omega^l \ (l =0,\dots,d-1)$. If they satisfy the conditions \eqref{SOSrelam}, then for any $n\neq d$ which is a divisor of $d$, 
\be \label{traceBm}
\Tr(A_{1,\alpha_1}^n)= 0 \qquad (\alpha=2,3).
\ee 
\end{lem}

\begin{proof}The proof is basically a repetition of a proof presented in Ref. \cite{Sarkar}.
It requires deriving two identities (whose details can be found in Appendix \ref{sec:selftesting})
that relate the traces of observables $A_{1,2}$ and $A_{1,3}$,
\begin{eqnarray}\label{Obs26m}
\Tr(A_{1,2}^{k})=\omega^{-k/m}\Tr(A_{1,3}^{k})
\end{eqnarray}
for $k = 1,\dots,\lfloor{d/2}\rfloor$ and,
\begin{eqnarray}\label{NewId1m}
\Tr(A_{1,2}^x)=\omega^{\frac{2tx}{m}}\,\Tr\left(A_{1,2}^{(2t+1)x}A_{1,3}^{-2tx}\right)
\end{eqnarray}
for $x = 1,\dots, \lfloor d/2\rfloor$ and any non-negative integer $t$. 
Now let us consider some non-negative integer $n\neq d$ which is a divisor of $d$, that is, $d/n \in \mathbb{N}$, which can be either even or odd. Note that any divisor of $d$ (except $d$ itself) is always smaller or equal to $d/2$. Whenever $d/n$ is even, that is, there exists some integer $t$ such that $n = d/2t$, we 
substitute $x = n = d/2t$ in Eq. \eqref{NewId1m}, which gives
\begin{eqnarray}\label{traces22}
\Tr(A_{1,2}^{n})=\omega^{d/m}\,\Tr\left(A_{1,2}^{d+n}A_{1,3}^{-d}\right).
\end{eqnarray}
Using the fact that $A_{1,\alpha_1}^d=\I$, the above relation simplifies to,
\be 
\Tr(A_{1,2}^{n})= %\mathrm{e}^{\frac{2\pi i}{m}}
\omega^{d/m}\Tr(A_{1,2}^{n}).
\ee 
As a consequence for $m\geq2$, we have that for any $n$ such that $d/n$ is even, $\Tr(A_{1,2}^{n})=0$. Using then Eq. \eqref{Obs26m}, one can similarly conclude that $\Tr(A_{1,3}^{n})=0$. 

Now, for any divisor $n$ of $d$ such that $d/n$ is odd, we choose $x = n = d/(2t+1)$ in Eq. \eqref{NewId1m}, which leads us to
\be \label{even-1}
\Tr(A_{1,2}^{n})= %\mathrm{e}^{\frac{2\pi i}{m}} 
\omega^{d/m}\w^{-n/m}\, \Tr(A_{1,3}^{n}).
\ee
Comparing the above expression with Eq. \eqref{Obs26m}, one directly concludes that $\Tr(A_{1,\alpha_1})=0$ for any $n$ such that $d/n$ is odd. Thus, 
we have shown that for any $n$ which is a divisor of $d$ such that $n\ne d$, $\Tr(A_{1,\alpha_1}^{n})=0$ for $\alpha_1=2,3$.
\end{proof}

In order to make use of the above lemma, in particular of Eq. (\ref{traceBm}), 
we need to recall another fact proven in Ref. \cite{Sarkar}.
\begin{lem}[\cite{Sarkar}] 
\label{le:pol}
Consider a real polynomial 
\begin{equation}\label{WX}
    W(x)=\sum_{i=0}^{d-1}\lambda_ix^i
\end{equation}
with rational coefficients $\lambda_i \in \Q$. Assume that $\omega^n$ with $\w=\mathrm{e}^{2\pi \mathbbm{i}/d}$ 
is a root of $W(x)$ for any $n$ being a proper divisor of $d$, i.e., $n\neq d$ such that 
$d/n\in \N$. Then, $\lambda_0=\lambda_1=\ldots=\lambda_{d-1}$.
\end{lem}

Now, denoting by $\lambda_{i,\alpha_1}$ the multiplicities of the eigenvalues $\omega^i$ $(i=1,\ldots,d-1)$ of the two observables $A_{1,2}$ and $A_{1,3}$, Eq. (\ref{traceBm}) implies that
\begin{eqnarray}\label{Vino}
\sum_{i=0}^{d-1}\lambda_{i,\alpha_1}\w^{ni}=0\qquad (\alpha_1=2,3),
\end{eqnarray}
where $n$ is a divisor of $d$ such that $n\ne d$. By virtue of Lemma \ref{le:pol} Eq. (\ref{Vino}) allows us to conclude that the multiplicities $\lambda_{i,\alpha_1}$ are all equal, or equivalently, $\lambda_{0,\alpha_1}=\ldots=\lambda_{d-1,\alpha_1}$. As a consequence, we have that
$\Tr(A_{1,\alpha_1}^{n})=0$ for all $n=1,2,\ldots,d-1$ and $\alpha_1=2,3$. 
Moreover, by employing the relation (\ref{SOSrelbm}) the latter fact can be directly extended to any observable
$A_{1,\alpha_1}$ measured by the first party. Precisely, taking trace of Eq. (\ref{SOSrelbm}) 
and using the explicit form of $R_{\alpha}^{(n)}$, one arrives at
\begin{equation}
\mu_{\alpha,n}^*\Tr\left(A_{1,2}^n\right)+\nu_{\alpha,n}^*\Tr\left(A_{1,\alpha+2}^n\right)+\tau_{\alpha,n}\Tr\left(A_{1,\alpha+3}^n\right)=0.
\end{equation}
Setting $\alpha=1$, one concludes from it that $\Tr (A_{1,4}^n)=0$. Then, setting $\alpha=2$ one obtains $\Tr (A_{1,5}^n)=0$. Continuing this procedure recursively for any $\alpha$ till $\alpha=m-2$, we conclude that $\Tr (A_{1,\alpha}^n)=0$ for all $n$ and $\alpha$. 
%\begin{proof}
%The proof of the above lemma can be found in .
%\end{proof}
%Without loss of generality, we take that the multiplicity of eigenvalue $\omega^i$ of $\mathcal{A}_{1,\alpha_1}$ is $\lambda_i$. From \textit{Lemma} \ref{le:traceless} which is stated above, we know that for all $n$ which is a divisor of $d$ and $n\neq d$, $\Tr(\mathcal{A}_{1,\alpha_1}^n)=0$ which implies $\sum^{d-1}_{i=0} \lambda_i(\omega^n)^i = 0.$ Now, using of \textit{Lemma} \ref{le:pol}, as stated above, we have that $\lambda_0=\dots=\lambda_{n-1}$ which allows us to infer that $\mathcal{A}_{1,\alpha_1} \in \C^d \otimes \mathcal{H}_{\mathcal{A}_{1}'}$.

The fact that the multiplicities of the eigenvalues of all the observables measured by the first party are equal means that they act on a Hilbert space whose dimension is a multiple of $d$, or, equivalently, that $\mathcal{H}_{1}=\mathbbm{C}^d\otimes \mathcal{H}_{1}'$ for some finite-dimensional Hilbert space $\mathcal{H}_{1}'$. Moreover, we have the following characterization of $A_{1,2}$ and $A_{1,3}$.

\begin{lem}\label{le:Fij}
Let us consider the unitary operators $A_{1,2}$ and $A_{1,3}$ acting on $\C^d\otimes\mathcal{H}_{1}'$ satisfying the conditions (\ref{SOSrelam}). Then, there exists a unitary $V_1:\mathcal{H}_{1}\rightarrow\mathcal{H}_{1}$ such that $V_1\,A_{1,2}\, V_1^\dagger = Z_d\otimes\mathbbm{1}_{1}'$ and $V_1\,A_{1,3}\,V_1^{\dagger}=T_{d,m}\otimes\mathbbm{1}_{1}'$ where $Z_d$ and $T_{d,m}$ are quantum observables defined in (\ref{ZdTd1}) and $\mathbbm{1}_{1}'$ is an identity acting on $\mathcal{H}_{1}'$.
\end{lem}

\begin{proof}
Due to the highly technical nature of the proof, it is deferred to Appendix \ref{sec:selftesting}. 
Here we only outline its major steps. 

First, the fact that, as proven above, the multiplicities of the eigenvalues of $A_{1,2}$
are equal one infers the existence of a unitary operation $V_1:\mathcal{H}_1\to\mathcal{H}_1$
for which
\begin{equation}
    V_1\,A_{1,2}\,V_1^{\dagger}=Z_d\otimes\mathbbm{1}_1'.
\end{equation}
Then, from the fact that $\mathcal{H}_{1}=\mathbbm{C}^d\otimes \mathcal{H}_{1}'$
one concludes that $V_1\,A_{1,3}\,V_1^{\dagger}$ can always be expressed in the following way
\be 
V_1\,A_{1,3}\,V_1^{\dagger}=\sum_{i,j=0}^{d-1}\ket{i}\!\bra{j}\otimes F_{ij}
\ee  
for some $F_{ij}$ acting on $\mathcal{H}_{{1}}'$. The final step, and in fact the most complicated one is to use the relations (\ref{SOSrelam}) to determine the form of the matrices $F_{ij}$, from which it follows that $V_1\,A_{1,3}\,V_1^{\dagger}=T_{d,m}\otimes\mathbbm{1}_{1}'$.
\end{proof}

We then show that the observables $A_{1,2}$ and $A_{1,3}$ are unitarily equivalent to the optimal measurements \eqref{measurementsm}. In fact, one can check that for the unitary matrix
$   W_{1}:\mathbbm{C}^d \rightarrow \mathbbm{C}^d $ given explicitly by
\begin{eqnarray}
W_1=\frac{1}{\sqrt{d}}\sum_{i,j=0}^{d-1}(-1)^{\delta_{j,0}}\omega^{-\frac{3i}{2m}+ij+\frac{j}{2}}\ket{i}\!\bra{j}.
\end{eqnarray}
the following relations hold true
\begin{equation}\label{derivedmeas}
W_{{1}}\,Z_d\,W_{{1}}^{\dagger}=\mathcal{O}_{1,2},\qquad
W_{{1}}\,T_{d,m}\,W_{{1}}^{\dagger}=\mathcal{O}_{1,3}.
\end{equation}
As a consequence, there exist a local unitary transformation
 $ U_{1}:\mathbbm{C}^d \otimes \mathcal{H}_{1}' \rightarrow \mathbbm{C}^d \otimes \mathcal{H}_{1}'$
such that 
\begin{eqnarray}\label{AiBimain2}
U_{1}\, A_{1,\alpha_1}\, U^{\dagger}_{1} = \mathcal{O}_{1,\alpha_1} \otimes \I_{1}'
\end{eqnarray} 
for $\alpha_1=2,3$. 

To find the other measurements of the first party $A_1$, we exploit the relation \eqref{SOSrelbm}. For this, we plug the obtained observables in \eqref{SOSrelbm} for $\alpha=1$, resulting in
\begin{equation}
U_1\,A_{1,4}\,U_1^{\dagger}=-\frac{1}{\tau_{1,1}}\left(\mu_{1,1}^*\,\mathcal{O}_{1,2} +\nu_{1,1}^*\,\mathcal{O}_{1,3}\right)\otimes\I_{1}'
\end{equation}
for all $n$. A key observation here is that the ideal observables (\ref{measurementsm}), (\ref{measurements2m}) and (\ref{measurements3m}) also satisfy the relations \eqref{SOSrelbm}. 
This directly implies that 
\begin{eqnarray}
U_1\,A_{1,4}\,U_1^{\dagger}=\mathcal{O}_{1,4}\otimes\I_{1}'.
\end{eqnarray}
Again, after exploiting \eqref{SOSrelbm} for $\alpha=2$,
\begin{equation}
U_1\,A_{1,5}\,U_1^{\dagger}=-\frac{1}{\tau_{1,2}}\left(\mu_{1,2}^*\mathcal{O}_{1,2} +\nu_{1,2}^*\mathcal{O}_{1,4}\right)\otimes\I_{1}',
\end{equation}
and using the relation (\ref{SOSrelbm}) for the ideal observables one has
\begin{eqnarray}
U_1\,A_{1,5}\,U_1^{\dagger}=\mathcal{O}_{1,5}\otimes\I_{1}'.
\end{eqnarray}
Applying this reasoning recursively for $\alpha=3,\ldots,m-2$, we conclude that
\begin{eqnarray}
U_{1} A_{1,\alpha_1} U^{\dagger}_{1} &=& \mathcal{O}_{1,\alpha_1} \otimes \I_{1}'  
\end{eqnarray}
for all $\alpha_1=1,\ldots,m$.

{\it{Observables of all the other parties.}} Following a similar strategy, we now show that for all the other parties the observables are equivalent to the optimal measurements \eqref{measurementsm}, \eqref{measurements2m} and (\ref{measurements3m}) up to some unitary transformation. To this end, we first find complementary SOS decompositions of the same Bell operator $\hat{\mathcal{I}}_{N,m,d}$ \eqref{eqASTAopm},
\begin{eqnarray}
\hspace{-0.1cm}\beta_Q\I-\mathcal{\hat{I}}_{N,m,d}&=&\frac{1}{2}\sum_{\alpha_1,\ldots,\alpha_N=1,2}\sum_{k=1}^{d-1}\left(P_{n,\alpha_1,\ldots,\alpha_N}^{(k)}\right)^{\dagger}P_{n,\alpha_1,\ldots,\alpha_N}^{(k)}\nonumber\\&&+\frac{m^{N-2}}{2}\sum_{\alpha=1}^{m-2}\sum_{k=1}^{d-1}\left(R_{n,\alpha}^{(k)}\right)^{\dagger}R_{n,\alpha}^{(k)},
\end{eqnarray}
where 
\begin{equation}
P_{n,\alpha_1,\ldots\alpha_N}^{(k)}=\I-A_{1,\alpha_1}^{(k)}\otimes\overline{A}_{n,\alpha_{n-1}+\alpha_n-1}^{(k)}\otimes\bigotimes_{\substack{i=2\\i\ne n}}^N A^{(-1)^{i-1}k}_{i,\alpha_{i-1}+\alpha_i-1}
\end{equation}
with $n=2,\ldots,N$ and for odd $n$,
\begin{equation}\label{SOS23}
\overline{A}_{n,\alpha_{n-1}+\alpha_n-1}^{(k)}=a_k A_{n,\alpha_{n-1}+\alpha_n-1}^{k}+a_k^*A_{n,\alpha_{n-1}+\alpha_n}^{k},
\end{equation}
whereas for even $n$,
\begin{equation}\label{SOS24}
\overline{A}_{n,\alpha_{n-1}+\alpha_n-1}^{(k)}=a_kA_{n,\alpha_{n-1}+\alpha_n-1}^{-k}+a_k^*A_{n,\alpha_{n-1}+\alpha_n-2}^{-k}.
\end{equation}
As before, in the above expressions we use the convention that $A_{n,\alpha+m}=\omega A_{n,\alpha}$ and $A_{n,0}=\omega^{-1}A_{n,m}$ for all $n,\alpha$.
%where $\mathcal{A}_{n,3}=\omega\mathcal{A}_{n,1}$ and $\mathcal{A}_{n,4}=\omega\mathcal{A}_{n,2}$.
Further,
\begin{eqnarray}\label{SOS25m}
R_{n,\alpha}^{(k)}=\mu_{\alpha,k}^* A_{n,2}^k+\nu_{\alpha,k}^*A_{n,\alpha+2}^k+\tau_{\alpha,k}A_{n,\alpha+3}^k
\end{eqnarray}
when $n$ is odd, and when $n$ is even
\begin{eqnarray}\label{SOS26m}
R_{n,\alpha}^{(k)}=\mu_{\alpha,k}A_{n,2}^{-k}+\nu_{\alpha,k}A_{n,\alpha+2}^{-k}+\tau_{\alpha,k}
A_{n,\alpha+3}^{-k}.
\end{eqnarray}

As in the previous case, the above SOS decompositions imply that the state $\ket{\psi_{N}}$ as well as the observables $A_{i,x_i}$ which maximally violate the Bell inequality (\ref{eqASTA})  satisfy 
\begin{equation}
    P_{n,\alpha_1,\ldots,\alpha_N}^{(k)}\ket{\psi_{N}}=0.
\end{equation}
As concluded before using \eqref{SOSrel2m}, we proceed in the similar way to obtain the relations for the observables of all the parties,
\begin{eqnarray}\label{SOSrelan}
\overline{A}_{n,\alpha_n}\overline{A}_{1,\alpha_n}^{\dagger}=\I, \quad  %\overline{\mathcal{A}}_{1,\alpha_1}^{(d)}=\I,\qquad
\text{and} \qquad \overline{A}_{n,\alpha_n}^{(k)}=\left[\overline{A}_{n,\alpha_n}^{(1)}\right]^k
\end{eqnarray}
for all $\alpha_n$ and $n=2,\ldots,N-1$. Also, using the fact that $a_{d-k}=a_{k}^{*}$ we have $\overline{A}_{n,\alpha_1}^{(d-k)}=\overline{A}_{n,\alpha_1}^{(k)\dagger}$ for 
any $k=1,\ldots,d-1$ and $\alpha_n=1,\ldots,m$. Furthermore, we have
\begin{eqnarray}
R_{n,\alpha}^{(k)}=0
\end{eqnarray}
for all $k=1,\ldots,d-1$ and $\alpha=1,2,\ldots,m-2$.
Note that for any observable $A_{n,\alpha}$ we obtained exactly the same relations 
as those derived previously for $A_{1,\alpha}$ given in Eq. \eqref{SOSrelam}. 
Consequently, we can straightforwardly conclude that for each 
party $A_n$ the corresponding Hilbert space 
decomposes as $\mathcal{H}_n=\mathbbm{C}^d\otimes\mathcal{H}_{n}'$ for some
finite-dimensional $\mathcal{H}_{n}'$, and, moreover, that there exists a unitary 
operation $V_n:\mathcal{H}_n\to \mathcal{H}_n$ such that
%
%$A_{n,\alpha_n}$ acts on Hilbert spaces of the form $\mathcal{H}_{\mathcal{A}_n}=\C^d \otimes %\mathcal{H}_{\mathcal{A}''_n}$ and
%
\begin{eqnarray}\label{derivedmeasn}
V_n\, A_{n,2}\, V_n^{\dagger}&=&Z_d\otimes\I_{n}', \nonumber\\ V_n\,A_{n,3}\,V_n^{\dagger}&=&T_{d,m}\otimes\I_{n}',
\end{eqnarray}
$\I_{i}'$ is the identity matrix acting on $\mathcal{H}_{n}'$ for any $n$. We now show that the obtained $Z_d$ and $T_{d,m}$ are equivalent to the ideal measurements \eqref{measurementsm}, \eqref{measurements2m} and \eqref{measurements3m} up to local unitary transformations
\begin{ufakt}
\label{fact:cglmp}
The unitary operators $W_2$, $W_{\mathrm{odd}}$ and $W_{\mathrm{ev}}$ acting on $\mathbbm{C}^d$  transform $Z_d$ and $T_{d,m}$ defined in Eq. \eqref{ZdTd1} to the ideal measurements given in Eqs. \eqref{measurementsm}, \eqref{measurements2m} and \eqref{measurements3m}, that is, one has for $x=2,3$,
\begin{equation}
 \mathcal{O}_{2,x}=W_2\,Z_d\,W_2^{\dagger}, \qquad 
\mathcal{O}_{2,3}=W_2\,T_{d,m}\,W_2^{\dagger},   
\end{equation}
for the second party $A_2$,
\begin{equation}
 \mathcal{O}_{\mathrm{odd},2}=W_{\mathrm{odd}}\,Z_d\,W_{\mathrm{odd}}^{\dagger},\qquad 
 \mathcal{O}_{\mathrm{odd},3}=W_{\mathrm{odd}}T_{d,m}W_{\mathrm{odd}}^{\dagger}   
\end{equation}
for all the parties numbered by odd numbers, and 
\begin{equation}
    \mathcal{O}_{\mathrm{odd},2}=W_{\mathrm{odd}}\,Z_d\,W_{\mathrm{odd}}^{\dagger},\qquad
    \mathcal{O}_{\mathrm{ev},3}=W_{\mathrm{ev}}\,T_{d,m}\,W_{\mathrm{ev}}^{\dagger}
\end{equation}
for the 'even' parties.
%
%\begin{equation}
% ,\qquad \mathcal{O}_{\mathrm{ev},3}=W_{\mathrm{ev}}T_{d,m}W_{\mathrm{ev}}^{\dagger}   
%\end{equation}
%
The unitary operators are given by
\begin{eqnarray}\label{Unitaries}
W_2&=&\frac{1}{\sqrt{d}}\sum_{i,j=0}^{d-1}(-1)^{\delta_{j,0}}\omega^{-\frac{2i}{m}+ij+\frac{j}{2}}\ket{d-1-i}\!\bra{j}, \nonumber\\
W_{\mathrm{odd}}&=&\frac{1}{\sqrt{d}}\sum_{i,j=0}^{d-1}(-1)^{\delta_{j,0}}\omega^{-\frac{i}{m}+ij+\frac{j}{2}}\ket{i}\!\bra{j},\nonumber\\
W_{\mathrm{ev}}&=&\frac{1}{\sqrt{d}}\sum_{i,j=0}^{d-1}(-1)^{\delta_{j,0}}\omega^{-\frac{i}{m}+ij+\frac{j}{2}}\ket{d-1-i}\!\bra{j}. 
\end{eqnarray}
\end{ufakt}
\begin{proof}
In Appendix \ref{sec:selftesting}, we show that with the suggested unitary transformations \eqref{Unitaries} the eigenvectors of $Z_d$ and $T_{d,m}$ are equivalent to the eigenvectors of the optimal measurements upto a phase of modulus 1.
\end{proof}
As a result, we conclude that there exist local unitary transformations
%
%\begin{equation}
 $   U_{i}:\mathbbm{C}^d \otimes \mathcal{H}_{i}' \rightarrow \mathbbm{C}^d \otimes \mathcal{H}_{i}'$
such that 
\begin{eqnarray}\label{AiBimain23}
U_{i}\, A_{i,\alpha_i}\, U^{\dagger}_{i} &=& \mathcal{O}_{i,\alpha_i} \otimes \I_{i}'  
\end{eqnarray}
for $\alpha_i=2,3$. Exploiting then the relations \eqref{SOS25m} and \eqref{SOS26m} one infers that \eqref{AiBimain23} holds true for all $\alpha_i=1,\ldots,m$. This concludes the part of the proof devoted to finding all the observables that violate the Bell inequality \eqref{eqASTA} maximally.

{\it{The state.}} Finally, using the derived optimal measurements and the relations \eqref{SOSrelam} we now show that the state which maximally violates the Bell inequality \eqref{eqASTA} is, up to local unitary transformations and some additional irrelevant degrees of freedom, the $N-$partite GHZ state \eqref{GHZstatem} of local dimension $d$.

%Let us first look at the quantity $\overline{\mathcal{A}}_{1,\alpha_1}^{(1)}$. Putting the observable from \eqref{Obs12}, we have
%\begin{eqnarray}
%\overline{\mathcal{A}}_{1,\alpha_1}^{(1)}&=&a_1\mathcal{A}_{1,\alpha_1}+a_1^*\mathcal{A}_{1,\alpha_1+1}\nonumber\\&=&\left(\sum_{i=0}^{d-2}\omega^{\zeta_m(\alpha_1)}\ket{i}\bra{i+1}+\omega^{(1-d)\zeta_m(\alpha_1)}\ket{d-1}\bra{0}\right)\otimes\I_{\mathcal{A}_n''}\nonumber\\
%\end{eqnarray}

To this aim, we exploit the fact that since $\ket{\psi_N}$ belongs to 
$\mathcal{H}_1\otimes\ldots\otimes\mathcal{H}_N$ with each local Hilbert space being
$\mathcal{H}_{n}=\mathbbm{C}^d\otimes\mathcal{H}_{n}'$ for any $n$, it can be decomposed as
\begin{eqnarray}\label{genstate}
\ket{\psi_{N}}=\sum_{i_1,\ldots, i_N=0}^{d-1}\ket{i_1,\ldots, i_N}\ket{\psi_{i_1,\ldots, i_N}},
\end{eqnarray}
where $\ket{\psi_{i_1,\ldots, i_N}}$ are some, in general unnormalized vectors from $\mathcal{H}_{1}'\otimes\ldots\otimes\mathcal{H}_N'$.

Now, considering the relations \eqref{SOSrelam}  for $k=1$ and and different values of 
$\alpha_i$'s we demonstrate that $\ket{\psi_{i_1,\ldots,i_N}}=0$ whenever there is a pair of indices $i_k\neq i_l$ for some $k\neq l$. Moreover, all components with $i_1=\ldots=i_N$ turn out to be equal. We thus find that the only state which maximally violates the Bell inequality \eqref{eqASTA} is given by
\begin{equation}
U_{1}\otimes\ldots\otimes U_{N}|\psi_{N}\rangle=\ket{\text{GHZ}_{N,d}} \otimes \ket{\mathrm{aux}_{N}},
\end{equation}
where $\ket{\mathrm{aux}_{N}}\in \mathcal{H}_{1}'\otimes\ldots \otimes \mathcal{H}_{N}'$. 
A more detailed explanation of this part of the proof can be found in Appendix B.
This completes the proof of our self-testing statement.
\end{proof}

%%%%%%%%%%%%%%%%%%%%%%%%%%%%%%%%%%%%%%%%%%%%%%%%%%%%%%%%%%%%%%%%

\section{Conclusions}

We proposed the first, to the best of our knowledge, self-testing statement for quantum states shared among arbitrary number of parties and of arbitrary local dimension that utilises a truly $d$-outcome Bell inequality. Contrary to the previous approach to self-testing of the GHZ states of Ref. \cite{Remik2}, which is a generalization of the results of Ref. \cite{Coladangelo}, our method does not rely on self-testing results for two-dimensional systems. Moreover, it allows for device-independent certification of the GHZ states of based on only two observables per observer, which is in fact the minimal number of observables allowing to observe quantum nonlocality and thus to make nontrivial self-testing statements. This lowers the experimental effort necessary to implement our scheme. Let us also notice that our self-testing method generalizes some previous results in a couple of ways. On one hand, we generalize the self-testing statement for two-qudit maximally entangled states derived in Ref. \cite{Sarkar} to an arbitrary number of observers as well as an arbitrary number of measurements. One the other hand, we generalize the self-testing statement based on the chained Bell inequalities given in Ref. \cite{chainedBell} to quantum systems of an arbitrary local dimension as well as an arbitrary number of parties.

%Moreover, we certify an arbitrary number of measurements per party which is a generalisation of the %self-testing results using the chained Bell inequality \cite{chainedBell} to arbitrary local %dimension and number of parties. As a consequence, this is the first instance where the multipartite %GHZ state of arbitrary local dimension could be certified using minimum number of measurements per %party, that is two, which is important for experimental scenarios as the number of correlations %which has to be measured in lab in order to certify the state is far less than any other schemes for %larger dimensions and parties. 

Our considerations provoke some further questions. First, as far as the possibility of experimental implementations of our results is concerned it would be interesting to study how robust is our self-testing statement to noises and experimental imperfections and how its robustness scales with the number of parties $N$. Deriving analytically such robust statements for any $d$ and $N$ is certainly a hard task, hence we leave it for future publications. Let us notice, nevertheless, 
that for the particular case of $N=m=2$ and $d=3$, 
the robustness of a self-testing statement for various two-qutrit entangled 
states based on violation of the Bell inequality (\ref{eqASTA}) and its variants
was investigated in Refs. \cite{Sarkar,Science} by using the numerical approach of Ref. \cite{SWAPtest}. Another route for future research would be to explore whether our self-testing scheme can be used for device-independent certification of randomness. In fact, it was shown 
in Ref. \cite{Sarkar} that in the bipartite case the maximal violation of the Bell inequality 
(\ref{eqASTA}) certifies $\log_2d$ bits of local randomness which by using the results of this work 
can be generalized to the GHZ states. It would be interesting to see whether our self-testing statement allows for certification of more randomness from these states
by taking into account measurements performed by groups of parties, not only a single one as in
Ref. \cite{Sarkar}. Finally, it would be interesting to see whether our scheme can be further generalized to obtain self-testing methods for other genuinely entangled multipartite states; in fact, no general scheme allowing to self-test any multipartite genuinely entangled state is known.

\textit{Acknowledgement.---} This work is supported by Foundation for Polish Science through the First Team project (No First TEAM/2017-4/31).

\onecolumngrid
\appendix
\section{Preliminaries}\label{sec:preliminaries}

In this appendix we collect all notions and facts necessary to prove our self-testing statement. First we recall the explicit form of 
the Bell operator \cite{ACTA} with arbitrary number of measurements per observer,
\begin{eqnarray}\label{eqASTAop}
\mathcal{\hat{I}}_{N,m,d}:&=&\sum_{\alpha_1,\ldots,\alpha_{N-1}=1}^m\sum_{k=1}^{d-1}\left(a_kA_{1,\alpha_1}^k+a_k^*A_{1,\alpha_1+1}^k\right)\otimes\bigotimes_{i=2}^NA^{(-1)^{i-1}k}_{i,\alpha_{i-1}+\alpha_i-1}
\end{eqnarray}
where,
\begin{equation}\label{app:ak}
     a_k=\frac{\omega^{\frac{2k-d}{4m}}}{2\cos(\pi/2m)}.
\end{equation}
and $A_{i,x_i}$ denotes the $x_i$th measurement of party $A_i$ with the additional condition that $\alpha_N=1$. Then, the maximal quantum value of $\langle\mathcal{\hat{I}}_{N,m,d}\rangle$ turns out to be $\beta_Q^d=m^N(d-1)$ and can be achieved by the $N$-qudit GHZ state, 
\begin{eqnarray}\label{GHZstate}
\ket{\text{GHZ}_{N,d}}=\frac{1}{\sqrt{d}}\sum_{i=0}^{d-1}\ket{i}^{\otimes N},
\end{eqnarray}
and observables given by 
\begin{equation}\label{measurements}
\mathcal{O}_{1,x}=U_{x}F_d\,\Omega_dF_d^{\dagger}U_{x}^{\dagger},\qquad
\mathcal{O}_{2,x}=V_{x}F_d^{\dagger}\Omega_dF_dV_{x}^{\dagger},
\end{equation}
for the first two parties, and
\begin{equation}\label{measurements2}
\mathcal{O}_{\mathrm{odd},x}=W_{x}F_d\,\Omega_dF_d^{\dagger}W_{x}^{\dagger}
\end{equation}
and
\begin{equation}\label{measurements3}
\mathcal{O}_{\mathrm{ev},x}=W_{x}^{\dagger}F_d^{\dagger}\Omega_dF_d W_{x}    
\end{equation}
for all other parties $A_i$ $(i=3,\ldots,N)$ numbered by odd and even numbers, respectively.
%\begin{eqnarray}\label{measurements2}
%\mathcal{O}_{3,x}&=&W_{x}F_d\,\Omega_dF_d^{\dagger}W_{x}^{\dagger}\nonumber\\    
%&\vdots&\nonumber\\
%\mathcal{O}_{N-1,x}&=&
%\left\{
%\begin{array}{ll}
%W_{x}F_d\Omega_dF_d^{\dagger}W_{x}^{\dagger}, & \,N\,\,\mathrm{even}\\[1ex]
%W_{x}^{\dagger}F_d^{\dagger}\Omega_dF_d W_{x}, & \,N\,\,\mathrm{odd}\\
%\end{array}
%\right.\nonumber\\
%\mathcal{O}_{N,x}&=&\left\{
%\begin{array}{ll}
%W_{x}^{\dagger}F_d^{\dagger}\Omega_dF_d W_{x}, & \,N\,\,\mathrm{even}\\[1ex]
%W_{x}F_d\Omega_dF_d^{\dagger}W_{x}^{\dagger}, & \,N\,\,\mathrm{odd}\\
%\end{array}
%\right.
%C_{z}=W_{z}F_d\Omega_dF_d^{\dagger}W_{z}^{\dagger}
%\end{eqnarray}
%
In the above formulas
\begin{equation}
F_d=\frac{1}{\sqrt{d}}\sum_{i,j=0}^{d-1}\omega^{ij}\ket{i}\!\bra{j},\qquad
\Omega_d=\mathrm{diag}[1,\omega,\ldots,\omega^{d-1}]
\end{equation}
with $\omega=\mathrm{exp}(2\pi\mathbbm{i}/d)$. Then, the unitary operations 
$U_x$, $V_x$ and $W_x$ are defined as
\begin{equation}
U_{x}=\sum_{j=0}^{d-1}\omega^{-j \gamma_{m}(x)}\proj{j},\qquad
V_x=\sum_{j=0}^{d-1}\omega^{j \zeta_{m}(x)}\proj{j},\qquad
W_x=\sum_{j=0}^{d-1}\omega^{-j \theta_{m}(x)}\proj{j},
\end{equation}
where
\begin{eqnarray}\label{measupara}
\gamma_m(x)=\frac{x}{m}-\frac{1}{2m},\qquad \zeta_m(x)=\frac{x}{m},\qquad \text{and}\qquad \theta_m(x)=\frac{x-1}{m}. 
\end{eqnarray}
The above observables can also be written in the matrix form as
\begin{eqnarray}\label{Obs12}
\mathcal{O}_{1,x}&=&\sum_{i=0}^{d-2}\omega^{\gamma_m(\alpha)}\ket{i}\!\bra{i+1}+\omega^{(1-d)\gamma_m(\alpha)}\ket{d-1}\!\bra{0},\nonumber\\  \mathcal{O}_{2,x}&=&\sum_{i=0}^{d-2}\omega^{\zeta_m(\alpha)}\ket{i+1}\!\bra{i}+\omega^{(1-d)\zeta_m(\alpha)}\ket{0}\!\bra{d-1}
\end{eqnarray}
for the first two parties, and 
\begin{eqnarray}\label{Obsoddeven}
\mathcal{O}_{\mathrm{odd},x}&=&\sum_{i=0}^{d-2}\omega^{\theta_m(\alpha)}\ket{i}\!\bra{i+1}+\omega^{(1-d)\theta_m(\alpha)}\ket{d-1}\!\bra{0}, \nonumber\\ \mathcal{O}_{\mathrm{ev},x}&=&\sum_{i=0}^{d-2}\omega^{\theta_m(\alpha)}\ket{i+1}\!\bra{i}+\omega^{(1-d)\theta_m(\alpha)}\ket{0}\!\bra{d-1}
\end{eqnarray}
for the remaining parties.

The sum-of-squares decomposition of the Bell operator $\mathcal{\hat{I}}_{N,m,d}$ found in Ref. \cite{ACTA} is given by
\begin{equation}\label{SOSAp}
    \beta_{Q}^d\mathbbm{1}-\mathcal{\hat{I}}_{N,m,d}=\frac{1}{2}\sum_{\alpha_1,\ldots,\alpha_{N-1}=1}^m\sum_{k=1}^{d-1}\left(P_{{\alpha_1,\ldots,\alpha_{N-1}}}^{(k)}\right)^{\dagger}P_{{\alpha_1,\ldots,\alpha_{N-1}}}^{(k)}+\frac{m^{N-2}}{2}\sum_{\alpha=1}^{m-2}\sum_{k=1}^{d-1}\left(R_{\alpha}^{(k)}\right)^{\dagger}R_{\alpha}^{(k)}
\end{equation}
with
\begin{equation}\label{SOSR1}
   P_{\alpha_1,\ldots,\alpha_{N-1}}^{(k)}=\I-\overline{A}_{1,\alpha_1}^{(k)}\otimes\bigotimes_{i=2}^NA^{(-1)^{i-1}k}_{i,\alpha_{i-1}+\alpha_i-1},\qquad \text{and}\qquad R_{\alpha}^{(k)}=\mu_{\alpha,k}^*A_{1,2}^k+\nu_{\alpha,k}^*A_{1,\alpha+2}^k+\tau_{\alpha,k}A_{1,\alpha+3}^k
\end{equation}
for $k=1,\ldots,d-1$ and all $\alpha_1,\ldots,\alpha_N$ and $\alpha\in\{1,2,\ldots,m-2\}$, where  
\begin{equation}
\overline{A}_{1,\alpha_1}^{(k)}=a_kA_{1,\alpha_1}^k+a_k^*A_{1,\alpha_1+1}^k.
\end{equation}
The coefficients $\mu_{\alpha,k}, \nu_{\alpha,k}$ and $\tau_{\alpha,k}$ are given by
\begin{eqnarray}\label{Rcoef1}
\mu_{\alpha,k}&=&\frac{\omega^{(\alpha+1)(d-2k)/2m}}{2\cos(\pi/2m)}\frac{\sin(\pi/m)}{\sqrt{\sin(\pi\alpha/m)\sin[\pi(\alpha+1)/m]}},\nonumber\\
\nu_{\alpha,k}&=&-\frac{\omega^{(d-2k)/2m}}{2\cos(\pi/2m)}\frac{\sqrt{\sin[\pi(\alpha+1)/m]}}{\sqrt{\sin(\pi\alpha/m)}},\nonumber\\
\tau_{\alpha,k}&=&\frac{1}{2\cos(\pi/2m)}\frac{\sqrt{\sin(\pi\alpha/m)}}{\sqrt{\sin[\pi(\alpha+1)/m]}}
\end{eqnarray}
for all $k$ and $\alpha=1,2,\dots,m-3$. For $\alpha=m-2$, we have
\begin{eqnarray}\label{Rcoef2}
\mu_{m-2,k}&=&-\frac{\w^{-k}\omega^{-(d-2k)/2m}}{2\cos(\pi/2m)\sqrt{2\cos(\pi/m)}},\nonumber\\
\nu_{m-2,k}&=&-\frac{\omega^{(d-2k)/2m}}{2\cos(\pi/2m)\sqrt{2\cos(\pi/m)}},\nonumber\\
\tau_{m-2,k}&=&\frac{\sqrt{2\cos(\pi/m)}}{2\cos(\pi/2m)}.
\end{eqnarray}
We finally introduce the following unitary matrices
\begin{equation}\label{ZdTd}
Z_d=\sum_{i=0}^{d-1}\omega^i\proj{i},\qquad T_{d,m}=\sum_{i=0}^{d-1}\omega^{i+\frac{1}{m}}\proj{i}-\frac{2\mathbbm{i}}{d}\sin\left(\frac{\pi}{m}\right)\sum_{i,j=0}^{d-1}(-1)^{\delta_{i,0}+\delta_{j,0}}\omega^{\frac{i+j}{2}-\frac{d-2}{2m}}|i\rangle\!\langle j|
\end{equation}
that are in fact also proper observables in our scenario.

%%%%%%%%%%%%%%%%%%%%%%%%%%%%%%%%%%%%%%%%%%%%%%%%%%%%%%%%%%%%
%%%%%%%%%%%%%%%%%%%%%%%%%%%%%%%%%%%%%%%%%%%%%%%%%%%%%%%%%%%%

\section{Proof of self-testing}\label{sec:selftesting}

Here we provide the full proof of theorem stated in the main text.
\begin{thm*}
Assume that the Bell inequality (\ref{eqASTAop}) is maximally violated by some state $\ket{\psi_{N}}\in\mathcal{H}_{1}\otimes\ldots\otimes\mathcal{H}_{N}$ and unitary observables $A_{i,\alpha}$ for all $i$ and $\alpha\in \{1,2,\ldots,m\}$. Then, there exist local unitary transformations
%
%\begin{equation}
 $   U_{i}: \mathcal{H}_{i} \rightarrow \mathbbm{C}^d \otimes \mathcal{H}_{i}'$
such that 
\begin{eqnarray}\label{AiBimain}
U_{i}\, A_{i,\alpha}\, U^{\dagger}_{i} = \mathcal{O}_{i,\alpha} \otimes \I_{i}'
\end{eqnarray}  
where $\I_{i}'$ are the identity matrices acting on $\mathcal{H}_{i}'$ for all $i's$ and the observables $\mathcal{O}_{i,\alpha}$ are defined in \eqref{Obs12} and \eqref{Obsoddeven}.
As a consequence, the state $\ket{\psi_{N}}$ which gives the maximal violation can be stated up to the freedom of local unitaries and the auxiliary state as
\be\label{state}
U_{1}\otimes \ldots \otimes U_{N}|\psi_{N}\rangle=\ket{\mathrm{GHZ}_{N,d}} \otimes \ket{\mathrm{aux}_{N}}, \ee 
where $\ket{\mathrm{aux}_{N}}\in \mathcal{H}_{1}'\otimes\ldots \otimes \mathcal{H}_{N}'$.
\end{thm*}
\begin{proof}The proof consists of two major steps.  In the first part of the proof, we begin by concentrating on the first party and
characterise its Hilbert space as well as observables; in fact, we show that 
in $\mathcal{H}_1$ one can identify a qudit Hilbert space $\mathbbm{C}^d$ and
prove the existence of a unitary operation that brings all $A_{1,x}$ to the ideal measurements
(\ref{measurements}). Then, we extend the above observations to the remaining parties. In the second part of the proof, we exploit the obtained observables to show that $\ket{\psi_N}$ is unitarily equivalent to the $N$-qudit GHZ state.

\textit{\textbf{The Hilbert space structure and characterization of observables.}} Before proceeding, without any loss of generality we can assume that the local reduced states $\rho_{i}$ corresponding to each party $A_i$ are full rank as the observables $A_{i,\alpha_i}$ can be certified only on the supports of the reduced states $\rho_{i}$ of $\ket{\psi_N}$. 

We begin by noting that the SOS decomposition (\ref{SOSAp}) it follows that the state $\ket{\psi_{N}}$ and the observables $A_{i,x_i}$ which maximally violate the above Bell inequality satisfy 
\begin{equation}
 P_{\alpha_1,\ldots,\alpha_N}^{(k)}\ket{\psi_{N}}=0,   
\end{equation}
which together with Eq. (\ref{SOSR1}) imposes that
\begin{eqnarray}\label{SOSrel2}
\overline{A}_{1,\alpha_1}^{(k)}\otimes\bigotimes_{i=2}^NA^{(-1)^{i-1}k}_{i,\alpha_{i-1}+\alpha_i-1}\ket{\psi_{N}}=\ket{\psi_{N}}\quad \forall k,\alpha_i.
\end{eqnarray}
for all $k$ and $\alpha_i$. Since $A_{i,\alpha_{i}}$ are unitary for all $i$ and $\alpha_i$, we can conclude that $\overline{A}_{1,\alpha_1}^{(k)}$ must obey the following conditions
\begin{eqnarray}\label{SOSrela}
\overline{A}_{1,\alpha_1}^{(k)}\overline{A}_{1,\alpha_1}^{(d-k)}=\I, \quad  %\overline{\mathcal{A}}_{1,\alpha_1}^{(d)}=\I,\qquad
\text{and} \quad \overline{A}_{1,\alpha_1}^{(k)}=\left[\overline{A}_{1,\alpha_1}^{(1)}\right]^k\quad \forall \alpha_1.
\end{eqnarray}
Note that $a_{d-k}=a_{k}^{*}$ and therefore $\overline{A}_{1,\alpha_1}^{(d-k)}=\overline{A}_{1,\alpha_1}^{(k)\dagger}$ for any $k=1,\ldots,d-1$. Additionally, we have another set of relations following from the SOS decomposition (\ref{SOSAp}), that is,
\begin{eqnarray}\label{SOSrelb}
R_{\alpha}^{(k)}\ket{\psi_{N}}=0\qquad \forall k,\alpha.
\end{eqnarray}
Due to the fact that $R_{\alpha}^{(k)}$ acts only on the first party's subsystem of $\ket{\psi_N}$, 
the above is equivalent to $R_{\alpha}^{(k)}\rho_{1}=0$, which taking into account the fact that $\rho_1$ is full rank implies
\begin{eqnarray}\label{SOSrelc}
R_{\alpha}^{(k)}=0
\end{eqnarray}
for all $k$ and $\alpha$. 

%
%
%
%\jk {Without loss of generality, we can assume that the support of second subsystem of %$\ket{\psi}$ is of full-rank which allows us to conclude (\ref{Ccond2}).}

In what follows we show that the conditions (\ref{SOSrela}) and (\ref{SOSrelb})
are enough to determine, up to local unitary operations, the observables $A_{1,x}$. First, in Lemma 
\ref{le:traceless} we show that the condition (\ref{SOSrela}) for $\alpha_1=2$ implies that 
\begin{equation}\label{eq17}
    \Tr(A_{1,2}^n)=\Tr(A_{1,3}^n)=0
\end{equation}
for any positive integer $n$ which is a divisor of $d$ such that $n\ne d$. Then Lemma \ref{le:pol} allows us to conclude \eqref{eq17} for all $n=1,2,\ldots,d-1$. Then using (\ref{SOSrelb}), one can simply conclude that $\Tr(A_{1,\alpha}^n)=0$ for all $\alpha\in\{1,2,\ldots,m\}$ and all $n$.
This allows us to conclude that the dimension of $A_{1}$'s Hilbert space $\mathcal{H}_{1}$ is $d\times D$ where $D$ is some arbitrary finite integer,
\begin{equation}
\mathcal{H}_{1}=\mathbb{C}^d\otimes\mathcal{H}_{1}',    
\end{equation}
where $\mathcal{H}_{1}'$ denotes space corresponding to the auxiliary degree of freedom. As a consequence, there exists a unitary matrix $V_{1}:\mathcal{H}_{1}\rightarrow \mathcal{H}_{1}$ such that 
\begin{equation}
    V_{1}\,A_{1,2}\,V_{1}^{\dagger}=Z_d\otimes\I_{1}',
\end{equation}
where $Z_d$ is defined in Eq. (\ref{ZdTd}). Using this equivalence, we find in Lemma \ref{le:Fij} that 
\begin{equation}
   V_{1} A_{1,3} V_{1}^{\dagger}=T_{d,m}\otimes \I_{1}'
\end{equation}
with $T_{d,m}$ defined in Eq. (\ref{ZdTd}). Next, we show that there exists unitary transformations $ U_{1}:\mathcal{H}_{1}\rightarrow \mathcal{H}_{1}$ such that 
\begin{eqnarray}
 U_{1}\,A_{1,\alpha_1}\,U_{1}^{\dagger}=\mathcal{O}_{1,\alpha_1}\otimes\I_{1}'\quad \text{for}\quad\alpha_1=2,3.
\end{eqnarray}
Finally, using the derived observables and \eqref{SOSrelb} we find the rest of the observables of $A_{1}$. We begin by proving lemma \eqref{le:traceless} which shows that the observables $A_{1,2}$ and $A_{1,3}$ are traceless.
%%%%%%%%%%
%%%%%%%%%%

\begin{lem}\label{le:traceless}
Consider two unitary observables $A_{1,\alpha_1}$  such that $\alpha_1= 2,3$ acting on a finite-dimensional Hilbert space whose eigenvalues are $\omega^l \ (l \in \{0,\dots,d-1\})$. If they they satisfy the conditions \eqref{SOSrela}, then for any $n\neq d$ which is a divisor of $d$, 
\be \label{traceB}
\Tr(A_{1,\alpha_1}^n)= 0.
\ee 
for $\alpha_1=2,3$.
\end{lem}
\begin{proof} First, we substitute the explicit forms of $\overline{A}_{1,2}$ and $a_k$ into both relations in \eqref{SOSrela} for $\alpha_1=2$, which after some algebra gives us two sets of equations with $k=1,\ldots,d-1$:
%
%\be 
%\left(a_kA_{1,2}^k+a_k^*A_{1,3}^k\right)  %\left(a_k^*A_{1,2}^{-k}+a_kA_{1,3}^{-k}\right) = \I.
%\ee
%
%Plugging in the value of $a_k$ from \eqref{app:ak} leads us to the %following condition,
%
\begin{equation}\label{Obs22}
\omega^{\frac{2k-d}{2m}}A_{1,2}^{k}A_{1,3}^{-k}+\omega^{-\frac{2k-d}{2m}}A_{1,3}^{k}A_{1,2}^{-k}=
2\cos\left(\frac{\pi}{m}\right)\mathbbm{1}
\end{equation} 
and 
%We then substitute $\overline{A}_{1,2}$ and $a_k$ into the second %relation in \eqref{SOSrela} for $\alpha_1=2$, which after some algebra %gives us 
%by again putting the explicit form of $\overline{A}_{1,2}$,
%
%\begin{eqnarray}
%a_{2k}A_{1,2}^{2k}+a_{2k}^*A_{1,3}^{2k} =\left(a_k %A_{1,2}^k+a_k^*A_{1,3}^k\right)\left(a_k A_{1,2}^k+a_k^*A_{1,3}^k\right) %\qquad k = 1,\dots, \left\lfloor{\frac{d}{2}}\right\rfloor.
%\end{eqnarray}
%
%A simple calculation using the explicit form of $a_k$ and some %trigonometric identities leads us to
%
\begin{eqnarray}\label{Obs23}
\omega^{k/m}A_{1,2}^{2k}+\omega^{-k/m}A_{1,3}^{2k}&=&A_{1,2}^kA_{1,3}^k+A_{1,3}^kA_{1,2}^{k},
\end{eqnarray}
Multiplying then Eq. \eqref{Obs23} by $A_{1,2}^{-k}$ and taking trace on both sides, one obtains
\begin{eqnarray}\label{Obs24}
\omega^{k/m}\Tr(A_{1,2}^{k})+\omega^{-k/m}\Tr(A_{1,3}^{2k}A_{1,2}^{-k})&=&2\Tr(A_{1,3}^k).
\end{eqnarray}
On the other hand, multiplying Eq. \eqref{Obs22} by $A_{1,3}^{k}$ and taking the trace on both sides, we get
\begin{eqnarray}\label{Obs25}
\omega^{\frac{2k-d}{2m}}\Tr(A_{1,2}^{k})+\omega^{-\frac{2k-d}{2m}}\Tr(A_{1,3}^{2k}A_{1,2}^{-k})=
2\cos\left(\frac{\pi}{m}\right)\Tr(A_{1,3}^{k}).
\end{eqnarray}
Then, one eliminates the term $\Tr(A_{1,3}^{2k}A_{1,2}^{-k})$ from \eqref{Obs24} and \eqref{Obs25} to arrive at 
\begin{eqnarray}\label{Obs26}
\Tr(A_{1,2}^{k})=2\omega^{-k/m}\frac{1-\cos(\pi/m)\omega^{-d/2m}}{1-\omega^{-d/m}}\Tr(A_{1,3}^{k})
\end{eqnarray}
which can be further simplified to
\begin{eqnarray}\label{Obs26}
\Tr(A_{1,2}^{k})=\omega^{-k/m}\Tr(A_{1,3}^{k})\qquad k = 1,\dots, \left\lfloor{\frac{d}{2}}\right\rfloor.
\end{eqnarray}
We have thus established a relation between traces of powers of the observables 
$A_{1,2}$ and $A_{1,3}$. It is thus enough to prove that they vanish for one of these observables. 
To this end, we prove the following observation.  
\begin{obs} \label{fact:4s}
The following identities hold true for any non-negative integer $t\in \mathbb{N} \cup \{0\}$ and $x=1,\ldots,\lfloor d/2\rfloor$:
\begin{equation}\label{NewId1}
    \Tr(A_{1,2}^x)=\omega^{\frac{2tx}{m}}\,\Tr\left(A_{1,2}^{(2t+1)x}A_{1,3}^{-2tx}\right).
    %\qquad x = 1,\dots, \left\lfloor{\frac{d}{2}}\right\rfloor.
\end{equation}
%
%and
%
%\begin{equation}\label{NewId2}
 %   \Tr(\mathcal{A}_{1,3}^t)=\omega^{\frac{2tx}{m}}\,\Tr\left(\mathcal{A}_{1,2}^{2ty}\mathcal{A}_{1,3}^{(-2t+1)y}\right)\qquad y = 1,\dots, \left\lfloor{\frac{d}{2}}\right\rfloor.
%\end{equation}
%
%\begin{equation}\label{obs311}
%\Tr(\mathcal{A}_{1,2}^xB^y_2)=\omega^{s(x+y)}\,\Tr\left(\mathcal{A}_{1,2}^{2s(x+y)+x}\mathcal{A}_{1,3}^{-2s(x+y)+y}\right).
%\end{equation}
\end{obs}
\begin{proof}
%We present the proof of the relation \eqref{NewId1} and the proof of the relation \eqref{NewId2} follows in the similar way. 
We prove this relation using mathematical induction. First, let us notice that it is trivially satisfied for $t=0$. Now, let us suppose that  \eqref{NewId1} holds true for $t=s-1$,
\begin{equation}\label{Obs31}
    \Tr(A_{1,2}^x)=\omega^{\frac{2(s-1)x}{m}}\,\Tr\left(A_{1,2}^{(2s-1)x}A_{1,3}^{-2(s-1)x}\right)\qquad x = 1,\dots, \left\lfloor{\frac{d}{2}}\right\rfloor.
\end{equation}
We will prove that the relation \eqref{NewId1} hold also true for $t=s$.
For this purpose, let us look at the right-hand side of Eq. (\ref{NewId1}) for $t=s$ and consider (\ref{Obs22}) for $k=2sx$, multiply it by $A_{1,2}^x$ and take the trace on both sides. This gives
\begin{equation}\label{Eq1}
   \omega^{\frac{4sx-d}{2m}} \Tr\left(A_{1,2}^{(2s+1)x}A_{1,3}^{-2sx}\right)+\omega^{\frac{d-4sx}{2m}}\Tr\left(A_{1,3}^{2sx}A_{1,2}^{(-2s+1)x}\right)=\cos\left(\frac{\pi}{m}\right)\Tr\left(A_{1,2}^x\right).
\end{equation}
We consider again Eq. (\ref{Obs22}) for $k=(2s-1)x$, 
multiply it by $A_{1,3}^x$ and then trace both sides, which results in
\begin{equation}\label{Eq2}
   \omega^{\frac{2(2s-1)x-d}{2m}} \Tr\left(A_{1,2}^{(2s-1)x}A_{1,3}^{-2(s-1)x}\right)+\omega^{\frac{d-2(2s-1)x}{2m}}\Tr\left(A_{1,3}^{2sx}A_{1,2}^{(-2s+1)x}\right)=\cos\left(\frac{\pi}{m}\right)\Tr\left(A_{1,3}^x\right),
\end{equation}
which after employing Eq. \eqref{Obs26} for $k=x$ simplifies to
\begin{eqnarray}\label{Eq2}
 \omega^{\frac{4(s-1)x-d}{2m}} \Tr\left(A_{1,2}^{(2s-1)x}A_{1,3}^{-2(s-1)x}\right)+\omega^{\frac{d-4sx}{2m}}\Tr\left(A_{1,3}^{2sx}A_{1,2}^{(-2s+1)x}\right)=\cos\left(\frac{\pi}{m}\right)\Tr\left(A_{1,2}^x\right).
\end{eqnarray}
Note that the above expression is valid only for $x = 1,\dots, \lfloor d/2\rfloor$. After subtracting Eq. (\ref{Eq2}) from Eq. (\ref{Eq1}) we arrive at
\begin{equation}\label{Eq3}
    \Tr\left(A_{1,2}^{(2s+1)x}A_{1,3}^{-2sx}\right)=\omega^{-\frac{2x}{m}}\left(A_{1,2}^{(2s-1)x}A_{1,3}^{-2(s-1)x}\right),
\end{equation}
which together with Eq. \eqref{Obs31} gives
\begin{equation}
    \Tr\left(A_{1,2}^{(2s+1)x}A_{1,3}^{-2sx}\right)=\omega^{-\frac{2sx}{m}}\Tr(A_{1,2}^x).
\end{equation}
This completes the proof of Observation \ref{fact:4s}.
\end{proof}

We are now in a position to prove Eq. \eqref{traceB}. 
Let $n$ be a divisor of $d$, that is, $d/n \in \mathbb{N}$. Note that any divisor of $d$ (except $d$ itself) is always smaller or equal to $d/2$. There are two possibilities of  
$d/n$ being even or odd. Whenever $d/n$ is even, that is, there exists some integer $t$ such that $n = d/2t$, we substitute $x = n = d/2t$ in Eq. \eqref{NewId1}, which gives
\begin{eqnarray}\label{traces22}
\Tr(A_{1,2}^{n})=\omega^{d/m}\,\Tr(A_{1,2}^{d+n}A_{1,3}^{-d}).
\end{eqnarray}
Using the fact that $A_{1,\alpha}^d=\I$, the above relation simplifies to
\be 
\Tr(A_{1,2}^{n})= \omega^{d/m}\Tr(A_{1,2}^{n}).
\ee 
%t
As a consequence, for any $m\geq2$, we have that for any $n$ such that $d/n$ is even, $\Tr(A_{1,2}^{n})=0$. Using then Eq. \eqref{Obs26} one can similarly conclude that $\Tr(A_{1,3}^{n})=0$. 

Now, for any divisor $n$ of $d$ such that $d/n$ is odd, we choose $x = n = d/(2t+1)$ in Eq. \eqref{NewId1}, which leads us to
\be \label{even-1}
\Tr(A_{1,2}^{n})= \w^{d/m} \w^{-n/m}\, \Tr\left(A_{1,3}^{n}\right).
\ee
Comparing the above expression with Eq. \eqref{Obs26}, one directly concludes that $\Tr(A_{1,\alpha})=0$ for any $n$ such that $d/n$ is odd 
and $n\leq d/2$. Thus, we have shown that for any $n$ which is a divisor of $d$, $\Tr(A_{1,\alpha}^{n})=0$ for $\alpha=2,3$. This completes the proof.
\end{proof}
%%%%%%%%%%%%%%%%

In order to exploit Eq. (\ref{traceB}) proven in the above lemma we need an important fact proven in Ref. \cite{Sarkar}.
\begin{lem}[\cite{Sarkar}] \label{le:pol}
Consider a real polynomial 
\begin{equation}\label{WX}
    W(x)=\sum_{i=0}^{d-1}\lambda_ix^i
\end{equation}
with rational coefficients $\lambda_i \in \Q$. Assume that $\omega^n$ with $\w=\mathrm{e}^{2\pi \mathbbm{i}/d}$ is a root of $W(x)$ for any $n$ being a proper divisor of $d$, i.e., $n\neq d$ such that $d/n\in \N$. Then, $\lambda_0=\lambda_1=\ldots=\lambda_{d-1}$.
\end{lem}

Using Lemma \ref{le:pol} one can show that if $\Tr(A_{1,\alpha}^{n})=0$ for $\alpha=2,3$ and $n$ being a divisor of $d$, the multiplicities of the eigenvalues of $A_{1,\alpha}$ are equal. As a consequence, we have that $\Tr(A_{1,\alpha}^{n})=0$ for all $n=1,2,\ldots,d-1$ and $\alpha=2,3$. 
Then, we consider the relation \eqref{SOSrelb} for $k=n$. After plugging the explicit form of $R_\alpha^{(n)}$ into it and taking trace, we obtain
\begin{eqnarray}
\mu_{\alpha,n}^*\Tr\left(A_{1,2}^n\right)+\nu_{\alpha,n}^*\Tr\left(A_{1,\alpha+2}^n\right)+\tau_{\alpha,n}\Tr\left(A_{1,\alpha+3}^n\right)=0.
\end{eqnarray}
Taking $\alpha=1$, one concludes that $\Tr(A_{1,4}^n)=0$. Then, taking $\alpha=2$, we obtain $\Tr(A_{1,5}^n)=0$. Continuing this procedure recursively for the remaining values of $\alpha$, we see that $\Tr(A_{1,\alpha}^n)=0$ for all $n$ and $\alpha$.

Now, we move onto finding the explicit form of the measurements $A_{1,\alpha}$ for $\alpha=2,3$.
%%%%%%%%%%%%%%%%%%%%%%%%%%%%%%%%
%LEMMA 3
%%%%%%%%%%%%%%%%%%%%%%%%%%%%%%%%

\begin{lem}\label{le:Fij}
Let us consider two unitary operators $A_{1,2}$ and $A_{1,3}$ acting on $\C^d\otimes \mathcal{H}_{1}'$ with eigenvalues $\w^l$ for $l=0,1,\ldots,d-1$ satisfying the conditions (\ref{SOSrela}).
%
%\begin{equation}\label{FijEq1}
%\mathcal{A}_{1,3}^k=-(k-1)\omega^{\frac{k}{2}}\mathcal{A}_{1,2}^k+\omega^{\frac{k-1}{2}}\sum_{m=0}^{k-1}\mathcal{A}_{1,2}^m\mathcal{A}_{1,3}\mathcal{A}_{1,2}^{k-%1-m}
%\end{equation}
%for all $k=1,\ldots,d$. 
Then, there exists a unitary $V_1:\mathcal{H}_{1}\rightarrow\mathcal{H}_{1}$ such that $V_1\,A_{1,2}\, V_1^\dagger = Z_d\otimes\mathbbm{1}_{1}'$ and 
$V_1\,A_{1,3}\,V_1^{\dagger}=T_{d,m}\otimes\mathbbm{1}_{{1}}'$ where $Z_d, T_{d,m}$ are defined in (\ref{ZdTd}).
\end{lem}
\begin{proof}
We begin by proving the following relation for $A_{1,2}$ and $A_{1,3}$:
\begin{equation}\label{FijEq1}
A_{1,3}^k=-(k-1)\omega^{\frac{k}{m}}A_{1,2}^k+\omega^{\frac{k-1}{m}}\sum_{t=0}^{k-1}A_{1,2}^tA_{1,3}A_{1,2}^{k-1-t} \qquad (k=1,\ldots,d).
\end{equation}
To this end, we use mathematical induction. First, it is not difficult to see that for $k=1$, the relation \eqref{FijEq1} is trivially satisfied as both its sides equal $A_{1,3}$. Assuming then that (\ref{FijEq1}) holds true for $k$, we will prove that it $k\to k+1$. With this aim, we consider \eqref{SOSrela} for $\alpha_1=2$ and rewrite it as
\begin{equation}
 \overline{A}_{1,2}^{(k+1)}=\overline{A}_{1,2}^{(k)}\overline{A}_{1,2}^{(1)} \qquad  (k=1,\dots,d-1).
\end{equation}
Plugging in the explicit form of $\overline{A}_{1,2}^{(k)}$ we arrive at
%a_{k+1}-a_{k}a_{1}=\omega^{\frac{k+1}{2m}}/2cos^2(\pi/2m)
\begin{eqnarray}
   A_{1,3}^{k+1}=-\omega^{\frac{k+1}{m}}A_{1,2}^{k+1}+\omega^{\frac{k}{m}}A_{1,2}^kA_{1,3}+\omega^{\frac{1}{m}}A_{1,3}^kA_{1,2},
\end{eqnarray}
which after substituting $A_{1,3}^k$ from Eq. \eqref{FijEq1} into it gives 
\begin{eqnarray}
   A_{1,3}^{k+1} &= &-\omega^{\frac{k+1}{m}}A_{1,2}^{k+1}+\omega^{\frac{k}{m}}A_{1,2}^kA_{1,3}+\omega^{\frac{1}{m}}\left[-(k-1)\omega^{\frac{k}{m}}A_{1,2}^k+\omega^{\frac{k-1}{m}}\sum_{t=0}^{k-1}A_{1,2}^tA_{1,3}A_{1,2}^{k-1-t}\right]A_{1,2}\nonumber\\
 &= &-k\omega^{\frac{k+1}{m}}A_{1,2}^{k+1}+\omega^{\frac{k}{m}}\sum_{t=0}^{k}A_{1,2}^tA_{1,3}A_{1,2}^{k-t}
\end{eqnarray} 

Now, from the fact that the multiplicities of all the eigenvalues of  $A_{1,\alpha}$ are equal, we conclude that there exists a unitary operation $V_1:\mathcal{H}_{1}\rightarrow\mathcal{H}_{1}$ such that $V_1\,A_{1,2}\,V_1^{\dagger}=Z_d\otimes\I_{1}'$. Moreover, we can always 
write $V_1\,A_{1,3}\, V_1^{\dagger}$ in the following way
\be  \label{B2form}
V_1\,A_{1,3}\,V_1^{\dagger}=\sum_{i,j=0}^{d-1}\ket{i}\!\bra{j}\otimes F_{ij},
\ee  
where $F_{ij}$ are some matrices acting on $\mathcal{H}_{1}'$. In order to make our further considerations simpler and easier to follow we drop the unitary $V_1$ acting on the observables for now and bring it back at the end of the proof; analogously, we write $\mathbbm{1}$ instead of $\mathbbm{1}_{1}'$. 

Our aim now is to determine $F_{ij}$ using relations (\ref{FijEq1}). First, we calculate $F_{ii}$ and then proceed to $F_{ij}$ for $i\neq j$. Eq. \eqref{FijEq1} for $k=d-1$ gives us
\begin{eqnarray}\label{FijEq3}
    A_{1,3}^\dagger=-(d-2)\omega^{\frac{d-1}{m}}A_{1,2}^{\dagger}+\omega^{\frac{d-2}{m}}\sum_{t=0}^{d-2}A_{1,2}^t A_{1,3}A_{1,2}^{d-t-2}.
\end{eqnarray}
Taking then $A_{1,2}=Z_d\otimes \I_{{1}}'$ and $A_{1,3}$ as given in Eq. (\ref{B2form}), the above expression \eqref{FijEq3} can be rewritten as
\begin{equation}\label{FijEq9}
 \sum\limits_{i,j=0}^{d-1}\ket{j}\!\bra{i}\otimes F_{ij}^\dagger =
-(d-2)\omega^{\frac{d-1}{m}}\sum_{i=0}^{d-1}\omega^{-i}\ket{i}\!\bra{i}\otimes \I + \omega^{\frac{d-2}{m}} \sum_{i,j=0}^{d-1}\sum_{t=0}^{d-2}\omega^{-2j+t(i-j)}\ket{i}\!\bra{j}\otimes F_{ij}.
\end{equation}
We can now project the first subsystem onto $|i\rangle\!\langle i|$ to obtain the following relation
\begin{eqnarray}\label{FijEq8}
  F_{ii}^\dagger =-(d-2)\omega^{\frac{d-1}{m}} \omega^{-i}\I +(d-1)\omega^{\frac{d-2}{m}} \omega^{-2i}F_{ii}.
\end{eqnarray}
Taking the conjugate of the above equation,
\begin{eqnarray}\label{FijEq85}
     F_{ii}=-(d-2)\omega^{-\frac{d-1}{m}}\omega^{i}\ \I +(d-1)\omega^{-\frac{d-2}{m}}\omega^{2i} F_{ii}^\dagger,
\end{eqnarray}
and substituting $F_{ii}^\dagger$ from Eq. \eqref{FijEq8} into it
\begin{equation} \label{Fijfact1}
F_{ii}=-(d-2)\omega^{-\frac{d-1}{m}}\omega^{i} \ \I -(d-2)(d-1)\omega^{\frac{1}{m}+i} \I+(d-1)^2F_{ii}, 
\end{equation}
we obtain 
\begin{equation}\label{TintoDeVerano}
    F_{ii}=\omega^{i+\frac{1}{m}} \left(\frac{d-1+\w^{-\frac{d}{m}}}{d}\right)\I=\omega^{i+\frac{1}{m}} \left(1-\frac{2\mathbbm{i}\sin(\pi/m)}{d}\w^{-\frac{d}{2m}}\right)\I.
\end{equation}

Now we focus on determining the matrices $F_{ij}$ for $i\neq j$. Our derivation is based on a sequence of observations. First, taking the $|j\rangle\!\langle i|$ elements of Eq. \eqref{FijEq9} for $i\neq j$, one finds the following equation
\begin{eqnarray}\label{fijEq12}
F_{ji}^\dagger=\omega^{\frac{d-2}{m}}\omega^{-2j}\sum_{t=0}^{d-2}\omega^{t(i-j)} F_{ij},
\end{eqnarray}
which after taking into account that $\sum_{t=0}^{d-2}\omega^{t(i-j)}=-\omega^{-(i-j)}$ for $i\ne j$, reduces to 
\be \label{fijEq12}
F_{ij}=-\w^{-\frac{d-2}{m}}\omega^{i+j}F_{ji}^\dagger.
\ee 
Note that \eqref{FijEq3} only relates the symmetric elements of $A_{1,3}$ in the form \eqref{fijEq12}. To find the explicit form of $F_{ij}$, we need to consider equations similar to 
Eq. (\ref{FijEq9}), however, with higher order terms in $F_{ij}$. To this end let us prove the following observation.

\begin{obs}\label{obs3.1}
The following conditions hold true for any $k=1,\ldots,d-1$ and $m\geq 2$,
 \begin{eqnarray} \label{FijObs3}
  -(k-1)\sum_{i,j=0}^{d-1}\omega^{ki}\ket{i}\!\bra{j}\otimes F_{ij}+\omega^{-\frac{1}{m}}\sum_{i,j=0}^{d-1}\ket{i}\!\bra{j}\otimes\left[\sum_{\substack{l=0\\l\ne i}}^{d-1}\left(\frac{\omega^{ki}-\omega^{kl}}{\omega^{i}-\omega^{l}}\right) F_{il}F_{lj}+k\omega^{(k-1)i}F_{ii}F_{ij}\right] \nonumber\\
  = -k\omega^{\frac{1}{m}}\sum_{i=0}^{d-1}\omega^{(k+1)i}\ket{i}\!\bra{i}\otimes \I+\sum_{i,j=0}^{d-1}\ket{i}\!\bra{j}\otimes\sum_{t=0}^{k}\omega^{k j+t(i-j)} F_{ij}.
  %
  %-(k-1)\omega^{\frac{k}{m}}\sum_{i,j=0}^{d-1}\omega^{ki}\ket{i}\!\bra{j}\otimes %F_{ij}+\omega^{\frac{k-1}{m}}\sum_{i,j=0}^{d-1}\ket{i}\!\bra{j}\otimes\left[\sum_{\substack{l=0\\l%\ne i}}^{d-1}\left(\frac{\omega^{ki}-\omega^{kl}}{\omega^{i}-\omega^{l}}\right) %F_{il}F_{lj}+k\omega^{(k-1)i}F_{ii}F_{ij}\right] \nonumber\\
  %= -k\omega^{\frac{k+1}{m}}\sum_{i=0}^{d-1}\omega^{(k+1)i}\ket{i}\!\bra{i}\otimes %\I+\omega^{\frac{k}{m}}\sum_{i,j=0}^{d-1}\left(\sum_{t=0}^{k}\omega^{k %j+t(i-j)}\ket{i}\!\bra{j}\otimes F_{ij}\right).
\end{eqnarray}
\end{obs}
\begin{proof}
Let us consider a trivial relation $A_{1,3}^{k+1}=A^k_{1,3}A_{1,3}$ and substitute in it 
$A_{1,3}^{k+1}$ and $A_{1,3}^k$ using Eq. (\ref{FijEq1}). This leads us to 
\begin{equation} \label{FijEq11a}
-k\w^{\frac{1}{m}}A_{1,2}^{k+1}+\sum_{t=0}^{k}A_{1,2}^tA_{1,3}A_{1,2}^{k-t}=-(k-1)A_{1,2}^kA_{1,3}+\omega^{-\frac{1}{m}}\sum_{t=0}^{k-1}A_{1,2}^tA_{1,3}A_{1,2}^{k-1-t}A_{1,3}.
\end{equation}
We now evaluate the sum appearing on the right-hand side by substituting the explicit forms of $A_{1,2}$ and $A_{1,3}$,
\begin{equation}
\sum_{t=0}^{k-1}A_{1,2}^tA_{1,3}A_{1,2}^{k-1-t}A_{1,3} 
= \sum_{i,j=0}^{d-1}\ket{i}\!\bra{j}\otimes\sum_{l=0}^{d-1}\sum_{t=0}^{k-1}\omega^{l(k-1)}\omega^{t(i-l)}F_{il}F_{lj}.
\end{equation}
Splitting the sum over $l$ into two parts: $l=i$ and $l\neq i$, 
and using the fact that 
\begin{equation}
    \sum_{t=0}^{k-1}\w^{t(i-l)}=\frac{1-\omega^{k(i-l)}}{1-\omega^{i-l}},
\end{equation}
we obtain
\begin{equation} \label{eq83}
\sum_{t=0}^{k-1}A_{1,2}^tA_{1,3}A_{1,2}^{k-1-t}A_{1,3}  
= \sum_{i,j=0}^{d-1}\ket{i}\!\bra{j}\otimes\left[\sum_{\substack{l=0\\l\ne i}}^{d-1}\left(\frac{\omega^{ki}-\omega^{kl}}{\omega^{i}-\omega^{l}}\right) F_{il}F_{lj}+k\omega^{(k-1)i}F_{ii}F_{ij}\right].
\end{equation}
Then, using similar arguments the sum on the left-hand side of Eq. (\ref{FijEq11a})
can be expressed as
\begin{equation}  \label{eq84}
\sum_{t=0}^{k}A_{1,2}^tA_{1,3}A_{1,2}^{k-t} = \sum_{i,j=0}^{d-1}\omega^{kj}\sum_{t=0}^{k}\omega^{t(i-j)}\ket{i}\!\bra{j}\otimes F_{ij}.
\end{equation} 
Plugging Eqs. (\ref{eq83}) and (\ref{eq84}) into Eq. (\ref{FijEq11a}) one arrives at 
(\ref{FijObs3}) which completes the proof of Observation \ref{obs3.1}.
\end{proof}

Equipped with the relation (\ref{FijObs3}) we can now proceed with the characterisation of $F_{ij}$. Precisely, the diagonal terms Eq. (\ref{FijObs3}) can be used to prove the following observation. 

\begin{obs}\label{Obs32}
The following conditions hold true for any pair $i\neq j$:
\begin{eqnarray}\label{FijEq25}
  F_{ij}F_{ij}^\dagger=\frac{4}{d^2}\sin^2\left(\frac{\pi}{m}\right)\I.
\end{eqnarray}
\end{obs}
\begin{proof}
Let us first consider Eq. (\ref{FijObs3}) and project the first subsystem onto $|i\rangle\!\langle i|$ which, after performing some simple algebra, gives
\begin{eqnarray}
 \sum_{\substack{l=0\\l\ne i}}^{d-1}\left(\frac{\omega^{ki}-\omega^{kl}}{\omega^{i}-\omega^{l}}\right) F_{il}F_{li} =
 k\omega^{ki}\left[2\omega^{\frac{1}{m}}F_{ii}-\omega^{-i}F_{ii}^2-\omega^{i+\frac{2}{m}}\I\right],
 %
 %2k \omega^{ki+\frac{1}{2}}F_{ii}-k\omega^{(k-1)i}F_{ii}^2-k\omega^{(k+1)i+1} \I_{\mathcal{A}_{1}''}.
\end{eqnarray}
This after substituting $F_{ii}$ from Eq. \eqref{TintoDeVerano} simplifies to 
\begin{equation}
 \sum_{\substack{l=0\\l\ne i}}^{d-1}\left(\frac{\omega^{ki}-\omega^{kl}}{\omega^{i}-\omega^{l}}\right) F_{il}F_{li} =-\frac{k}{d^2}\omega^{i(k+1)+\frac{2}{m}}(1-\w^{-d/m})^2\I,
 \end{equation}
and then, after a few algebraic manipulations, to 
\begin{equation}
\sum_{\substack{l=0\\l\ne i}}^{d-1}\left(\frac{1-\omega^{k(l-i)}}{1-\omega^{i-l}}\right)F_{il}F_{li}\omega^{-\left(i+l+\frac{2}{m}\right)}\w^{\frac{d}{m}}=\frac{k}{d^2}\w^{\frac{d}{m}}(1-\w^{-d/m})^2 \I.
\end{equation}
Then, after taking into account \eqref{fijEq12}, changing the index $l$ to $j$, and simplifying the right hand side we have 
\begin{eqnarray}\label{FijFact2}
\sum_{\substack{j=0\\j\ne i}}^{d-1}\left(\frac{1-\omega^{k(j-i)}}{1-\omega^{i-j}}\right)F_{ij}F_{ij}^{\dagger}=\frac{4k}{d^2}\sin^2\left(\frac{\pi}{m}\right) \I,
\end{eqnarray}
for $k=0,\ldots,d-1$ and $i=0,\ldots,d-1$. 
We then multiply the above expression (\ref{FijFact2}) by $\omega^{kn}$ with $k=0,\ldots,d-1$ and
$n=1,\ldots,d-1$ and then sum the resulting relation over all $k'$s,
\begin{equation}\label{formulaTwo}
-\sum_{\substack{j=0\\j\ne i}}^{d-1}\frac{1}{1-\omega^{i-j}}F_{ij}F_{ij}^\dagger\sum_{k=0}^{d-1}\omega^{k(j-i+n)}=\frac{4}{d^2}\sin^2\left(\frac{\pi}{m}\right)\sum_{k=0}^{d-1}k\omega^{kn}\I .
\end{equation}
Exploiting the following identities 
\begin{eqnarray} \label{FijIden2} 
 \sum_{k=0}^{d-1}\omega^{kn}=0,\quad \sum_{k=0}^{d-1}k\omega^{kn}=\frac{d}{\omega^n-1},\quad \text{and}\quad \sum_{k=0}^{d-1}\omega^{k(j-i+n)}=\delta_{j,i-n \mod d}
\end{eqnarray}
that are satisfied for any $n=1,\ldots,d-1$, we arrive at the following relation
\begin{eqnarray}
F_{i(i-n \mod d)}F_{i(i-n\mod d)}^\dagger=\frac{4}{d^2}\sin^2\left(\frac{\pi}{m}\right)\I.
\end{eqnarray}
Note that for any 
$i=0,\ldots,d-1$ there exist $n=1,\ldots,d-1$ such that $i-n\mod d$ is any number from $\{0,\ldots,d-1\}$ and is different than $i$. This completes the proof of Observation \ref{Obs32}.
\end{proof}

While \eqref{FijEq25} tell us a lot about the matrices $F_{ij}$ it is still not enough to 
determine their explicit form. To complete the characterisation
we consider a unitary operation $\widetilde{V}:\mathcal{H}_1\to\mathcal{H}_1$ 
of the following form 
\begin{eqnarray}
\widetilde{V} = \sum^{d-1}_{i=0} \ket{i}\!\bra{i} \otimes \widetilde{V}_i, 
\end{eqnarray} 
where $\widetilde{V}_i:\mathcal{H}_1'\to \mathcal{H}_1'$ are unitary operations defined as
%(\textbf{Remik: There should be minus in $\widetilde{U}_i$, right?}-Yes, correct!)
%
\begin{equation}
    \widetilde{V}_0=\I,\qquad \widetilde{V}_i=-\frac{d\mathbbm{i}}{2\sin(\pi/m)}\omega^{-\frac{i}{2}+\frac{d-2}{2m}}F_{0i} \qquad (i=1,\ldots,d-1).
\end{equation}
Importantly, $\widetilde{V}$ commutes with $Z_d$ and thus preserves the form of $A_{1,2}$.

We then have 
\begin{eqnarray}
 \widetilde{V}\,A_{1,3}\,\widetilde{V}^\dagger  = \sum^{d-1}_{i,j=0} \ket{i}\!\bra{j} \otimes \tilde{F}_{ij},
\end{eqnarray}
where we denoted $\tilde{F}_{ij}=\widetilde{V}_i\,F_{ij}\,\widetilde{V}_j^\dagger$.
For a remark, all the algebraic relations for $F_{ij}$ obtained so far hold true for $\tilde{F}_{ij}$, and $\tilde{F}_{ii} = F_{ii}$.
Now, we see that
\begin{eqnarray}\label{FijEq28}
  \tilde{F}_{0j}=\widetilde{V}_0\,F_{0j}\,\widetilde{V}_j^\dagger
  %=\frac{d}{2}\omega^{\frac{2j+d}{4}+\frac{2-d}{2m}}F_{0j}F_{0j}^\dagger
  =\frac{2\mathbbm{i}}{d}\sin\left(\frac{\pi}{m}\right)\omega^{\frac{j}{2}+\frac{2-d}{2m}}\I,
\end{eqnarray} 
where we employed Eq. (\ref{FijEq25}) for $i=0$.
Now, using Eq. (\ref{fijEq12}) we obtain that $\tilde{F}_{j0}=\tilde{F}_{0j}$.

In the remaining part of the proof of Lemma \ref{le:Fij}, we focus on the elements $F_{ij}$ for $i,j \neq 0$ and  $i\neq j$. To do so, we exploit the off-diagonal elements of \eqref{FijFact2}.
\begin{obs}\label{Obs33}
The following conditions hold true
\begin{equation} \label{FijFact3}
\sum_{\substack{i=1\\i\ne j}}^{d-1}\left(\frac{1-\omega^{ki}}{1-\omega^{i}}\right) \omega^{\frac{i}{2}}F_{ij}=
\frac{1-\w^{-d/m}}{d}\omega^{\frac{j}{2}+\frac{1}{m}}\left(k+\frac{1-\omega^{kj}}{1-\omega^{j}}\omega^{j}\right) \I, \qquad k,j=1,\ldots,d-1.
\end{equation}
\end{obs}
\begin{proof}
Taking the inner product with $\bra{i} \ . \ |j\rangle$ (where $i\neq j$) on the both side of \eqref{FijObs3} we obtain
\begin{eqnarray}
 -(k-1)\omega^{ki}F_{ij}+\omega^{-\frac{1}{m}}\sum_{\substack{l=0\\l\ne i}}^{d-1}\left(\frac{\omega^{ki}-\omega^{kl}}{\omega^{i}-\omega^{l}}\right) F_{il}F_{lj}+k\omega^{(k-1)i}\omega^{-\frac{1}{m}}F_{ii}F_{ij}=\frac{\omega^{(k+1)i}-\omega^{(k+1)j}}{\omega^{i}-\omega^{j}}F_{ij}. 
\een
Rearranging some terms and using $F_{ii}$ from \eqref{TintoDeVerano}, we have
 \ben 
 \sum_{\substack{l=0\\l\ne i}}^{d-1}\frac{\omega^{ki}-\omega^{kl}}{\omega^{i}-\omega^{l}} F_{il}F_{lj}=\omega^{\frac{1}{m}}\left[\frac{\omega^{(k+1)i}-\omega^{(k+1)j}}{\omega^{i}-\omega^{j}}+\left(\frac{(1-\w^{-d/m})k}{d}-1\right)\omega^{ki}\right]F_{ij}.
\end{eqnarray}
Next, we set $i=0$ and obtain
\begin{eqnarray}
\sum_{l=1}^{d-1}\frac{1-\omega^{kl}}{1-\omega^{l}} F_{0l}F_{lj}=\omega^{\frac{1}{m}}\left(\frac{1-\omega^{(k+1)j}}{1-\omega^{j}}+\frac{(1-\w^{-d/m})k}{d}-1\right)F_{0j}.
\end{eqnarray}
Substituting ${F}_{0j}$ from \eqref{FijEq28},
\begin{equation}\label{VihnoVerde}
    \sum_{l=1}^{d-1}\frac{1-\omega^{kl}}{1-\omega^{l}}\omega^{\frac{l}{2}} F_{lj}=\omega^{\frac{j}{2}+\frac{1}{m}}\left(\frac{1-\omega^{(k+1)j}}{1-\omega^{j}}+\frac{(1-\w^{-d/m})k}{d}-1\right)\I.
\end{equation}
Taking the term corresponding to $l=j$ out of the sum and expressing $F_{jj}$ with the aid of Eq. \eqref{TintoDeVerano} we arrive at the desired formula Eq. (\ref{FijFact3}), which completes the proof of Observation \ref{Obs33}.
%
%\begin{equation}
%\sum_{\substack{l=1\\l\ne j}}^{d-1}\left(\frac{1-\omega^{kl}}{1-\omega^{l}}\right) %\omega^{\frac{l}{2}}F_{lj}=
%\frac{(1-\w^{-d/m})}{d}\omega^{\frac{j}{2}+\frac{1}{m}}\left(k+\frac{1-\omega^{kj}}{1-\omega^{j}}\om%ega^{j}\right) \I.
%\end{equation}
%
%Finally, changing the index $l$ to $i$ leads us to \eqref{FijFact3}, completing the proof of %Observation (\ref{Obs33}).
%
\end{proof}
We can finally find $F_{ij}$ for $i\ne j$ and $i,j\ne 0$. For this, we multiply Eq. \eqref{FijFact3} by $\omega^{-kn}$ with $n=1,\ldots,d-1$ such that $n\neq j$ and then 
sum both sides of the resulting formula over $k=0,\ldots,d-1$, obtaining
\begin{equation}
\sum_{\substack{i=1\\i\ne j}}^{d-1}\frac{\omega^{i/2}}{1-\omega^{i}}F_{ij}\sum_{k=0}^{d-1}\left(\omega^{-kn}-\omega^{k(i-n)}\right)=
\frac{(1-\w^{-d/m})}{d}\omega^{\frac{j}{2}+\frac{1}{m}}\left[\sum_{k=0}^{d-1}k\omega^{-kn}+\frac{\omega^{j}}{1-\omega^{j}}
\sum_{k=0}^{d-1}\left(\omega^{-kn}-\omega^{k(j-n)}\right)\right]\I.
\end{equation}
Notice that in the above equation the first sum over $k$ on the left-hand side as well as
the last two sums on the right-hand side simply vanish for $n\neq j$. Now, exploiting Eq. (\ref{FijIden2}) as well as the fact that $\sum_{k=0}^{d-1}\omega^{k(n-i)}=d\delta_{n,i}$, 
and the identity
\begin{eqnarray} \label{FijIden2} 
  \sum_{k=0}^{d-1}k\omega^{kn}=\frac{d}{\omega^n-1}, \qquad n=1,\ldots,d-1.
\end{eqnarray}
proven in \cite{Sarkar}, we obtain
%
%Note that the above equation trivially holds for $k=0$. Employing the facts $\sum_{k=0}^{d-1}\omega^{kn}=0$ if $n\ne0$, $\sum_{k=0}^{d-1}\omega^{k\alpha}=d$ if %$\alpha=0$, and \eqref{FijIden2}, we get
%
\begin{equation}
-d\frac{\omega^{n/2}}{1-\omega^{n}}F_{nj}=\frac{(1-\w^{-d/m})}{d}\omega^{\frac{j}{2}+\frac{1}{m}}\left(\frac{d}{\omega^{-n}-1}\right)\I,
\end{equation}
which after simple algebra leads us to,
\begin{eqnarray}\label{FijEq34}
F_{ij}=-\frac{(1-\w^{-d/m})}{d}\omega^{\frac{i+j}{2}+\frac{1}{m}}\I=
-\frac{2\mathbbm{i}\sin(\pi/m)}{d}\omega^{\frac{i+j}{2}+\frac{2-d}{2m}}\I, \qquad i, j=1,\ldots,d-1,\ \ \  i\ne j .
\end{eqnarray}
Finally, taking into account Eqs. (\ref{B2form}), \eqref{TintoDeVerano}, \eqref{FijEq28} and 
\eqref{FijEq34} we conclude that there exists a unitary operation $U_1=\widetilde{V}\overline{V}_1$ such that $U_1\,A_{1,2}\,U_1^{\dagger}=Z_d\otimes\mathbbm{1}$ and
\be 
U_1\, A_{1,3}\, U_1^\dagger = T_{d,m} \otimes \I
\ee 
with $T_{d,m}$ given by
\begin{eqnarray}\label{Td}
T_{d,m}=\sum_{i=0}^{d-1}\omega^{i+\frac{1}{m}}\proj{i}-\frac{2\mathbbm{i}}{d}\sin\left(\frac{\pi}{m}\right)\sum_{i,j=0}^{d-1}(-1)^{\delta_{i,0}+\delta_{j,0}}\omega^{\frac{i+j}{2}-\frac{d-2}{2m}}|i\rangle\!\langle j|
\end{eqnarray}
This completes the characterisation of $A_{1,2}$ and $A_{1,3}$.
\end{proof}

%%%%%%%%%%%%%%%%%%%%%%%%
%FINAL OBSERVATION
%%%%%%%%%%%%%%%%%%%%%%%%
%%%%%%%%%%%%%%%%%%%%%%%%%%%%%%%%%%%%%%%%%%%%%%%%%%%%%
%
Before deriving all the other observables of $A_1$, we find the observables $A_{i,2}$ and  $A_{i,3}$ for all $i$. Then, we show that the obtained measurements are unitarily equivalent to optimal measurements $\mathcal{O}_{i,2}$ and $\mathcal{O}_{i,3}$ for all $i$ \eqref{Obs12} and \eqref{Obsoddeven}. For this, we find different sum of squares decomposition of the Bell operator \eqref{eqASTAop}. Precisely, the Bell operator \eqref{eqASTAop} can also be written using the following sum of squares decomposition,
\begin{eqnarray}
\beta_Q\I-\mathcal{\hat{I}}_{N,m,d}=\frac{1}{2}\sum_{\alpha_1,\ldots\alpha_N=1,2}\sum_{k=1}^{d-1}P_{n,\alpha_1,\ldots\alpha_N}^{(k)}P_{n,\alpha_1,\ldots\alpha_N}^{(k)\dagger}+\frac{m^{N-2}}{2}\sum_{\alpha=1}^{m-2}\sum_{k=1}^{d-1}\left(R_{n,\alpha}^{(k)}\right)^{\dagger}R_{n,\alpha}^{(k)}
\end{eqnarray}
where $n\in\{2,\ldots,N\}$ and,
\begin{eqnarray}
P_{n,\alpha_1,\ldots,\alpha_N}^{(k)}=\I-A_{1,\alpha_1}^{(k)}\otimes\overline{A}_{n,\alpha_{n-1}+\alpha_n-1}^{(k)}\otimes\bigotimes_{\substack{i=2\\i\ne n}}^N A^{(-1)^{i-1}k}_{i,\alpha_{i-1}+\alpha_i-1}
\end{eqnarray}
such that when $n$ is odd,
\begin{eqnarray}\label{SOS23}
\overline{A}_{n,\alpha_{n-1}+\alpha_n-1}^{(k)}=a_kA_{n,\alpha_{n-1}+\alpha_n-1}^{k}+a_k^*A_{n,\alpha_{n-1}+\alpha_n}^{k}
\end{eqnarray}
and when $n$ is even,
\begin{eqnarray}\label{SOS24}
\overline{A}_{n,\alpha_{n-1}+\alpha_n-1}^{(k)}=a_kA_{n,\alpha_{n-1}+\alpha_n-1}^{-k}+a_k^*A_{n,\alpha_{n-1}+\alpha_n-2}^{-k}
\end{eqnarray}
such that $A_{n,\alpha+m}=\omega A_{n,\alpha}$ and $A_{n,0}=\omega^{-1}A_{n,m}$ for all $n,\alpha$.
%where $\mathcal{A}_{n,3}=\omega\mathcal{A}_{n,1}$ and $\mathcal{A}_{n,4}=\omega\mathcal{A}_{n,2}$.
Further,
\begin{eqnarray}\label{SOS25}
R_{n,\alpha}^{(k)}=\mu_{\alpha,k}^*A_{n,2}^k+\nu_{\alpha,k}^*A_{n,\alpha+2}^k+\tau_{\alpha,k}A_{n,\alpha+3}^k
\end{eqnarray}
when $n$ is odd, and when $n$ is even
\begin{eqnarray}\label{SOS26}
R_{n,\alpha}^{(k)}=\mu_{\alpha,k}A_{n,2}^{-k}+\nu_{\alpha,k}A_{n,\alpha+2}^{-k}+\tau_{\alpha,k}A_{n,\alpha+3}^{-k}
\end{eqnarray}
where the coefficients $\mu_{\alpha,k}$, $\nu_{\alpha,k}$ and $\tau_{\alpha,k}$ are given in Eq. \eqref{Rcoef1} and Eq. \eqref{Rcoef2}. 

It is important to note that the above sum-of-squares decompositions can be used to obtain exactly the same relations for $\overline{A}_{n,\alpha_{n-1}+\alpha_n-1}$ as those in Eqs. \eqref{SOSrela} and \eqref{SOSrelb}, that is,
\begin{eqnarray}\label{SOSrela1}
\overline{A}_{n,\alpha_{n-1}+\alpha_n-1}\overline{A}_{n,\alpha_{n-1}+\alpha_n-1}^{\dagger}=\I, \quad
\text{and} \quad \overline{A}_{n,\alpha_{n-1}+\alpha_n-1}^{(k)}=\left[\overline{A}_{n,\alpha_{n-1}+\alpha_n-1}^{(1)}\right]^k\quad \forall \alpha_{n-1},\alpha_n
\end{eqnarray}
for any $\alpha_{n-1}$ and $\alpha_n$ and any $n=2,\ldots,N$. Moreover, the above SOS decompositions
imply also that
\begin{eqnarray}\label{SOSrelb1}
R_{n,\alpha}^{(k)}=0
%\quad \forall k,\quad \alpha=1,2,\ldots,m-2.
\end{eqnarray}
for any $k$ and $\alpha=1,\ldots,m-2$.

It is important to observe that the form of $ \overline{A}_{n,\alpha_{n-1}+\alpha_n-1}^{(k)}$ in \eqref{SOS23} and \eqref{SOS24} are exactly same as $\overline{A}_{1,\alpha_1}$ and the form of $R_{n,\alpha}^{(k)}$ in \eqref{SOS25} and \eqref{SOS26} is exactly similar as $R_{1,\alpha}^{(k)}$. Thus, the technique to find observables $A_{n,\alpha}$ is exactly same for all $n,\alpha$. 

As a consequence, we consider Eq. \eqref{SOSrela1} and Eq. \eqref{SOSrelb1} for $\alpha_n=1$ and $\alpha_{n-1}=2$ and following the similar techniques as done above to derive $A_{1,2}$ and $A_{1,3}$, we conclude that $U\,A_{n,2}\,U^{\dagger}=Z_d\otimes\I_{n}'$ and $U\,A_{n,3}\,U^{\dagger}=T_{d,m}\otimes\I_{n}'$ where $Z_d$ and $T_{d,m}$ are given in \eqref{ZdTd}. Finally, we show that $U_{i}\,A_{n,2}\,U_{i}^{\dagger}=\mathcal{O}_{n,2}\otimes\I_{n}'$ and $U_{i}\,A_{n,3}\,U_{i}^{\dagger}=\mathcal{O}_{n,3}\otimes\I_{n}'$ where $\mathcal{O}_{n,2}$ and $\mathcal{O}_{n,3}$ are the optimal measurements presented in Eq. \eqref{Obs12} and Eq. \eqref{Obsoddeven}. For this, we first express the ideal measurements using their eigen-decomposition,
\begin{equation}
 \mathcal{O}_{n,x}=\sum^{d-1}_{r=0} \w^r \ket{r}\!\bra{r}_{n,x}
\end{equation}
with $x=2,3$ and $n=1,2,\ldots,N$, where the eigenvectors are defined as
\begin{eqnarray}\label{idealeigen}
 \ket{r}_{1,x}&=&\frac{1}{\sqrt{d}}\sum_{q=0}^{d-1} \omega^{(r-\gamma_m(x))q}\ket{q}, \nonumber\\
 \ket{r}_{2,x}&=&\frac{1}{\sqrt{d}}\sum_{q=0}^{d-1} \omega^{-(r-\zeta_m(x))q}\ket{q},\nonumber\\
 \ket{r}_{n_{\mathrm{odd}},x}&=&\frac{1}{\sqrt{d}}\sum_{q=0}^{d-1} \omega^{(r-\theta_m(x))q}\ket{q},\nonumber\\
 \ket{r}_{n_{\mathrm{ev}},x}&=&\frac{1}{\sqrt{d}}\sum_{q=0}^{d-1} \omega^{-(r-\theta_m(x))q}\ket{q}
 \end{eqnarray}
where $\gamma_m(x)$, $\zeta_m(x)$ and $\theta_m(x)$ are given in Eq. \eqref{measupara} and $\{\ket{q}\}$ is the computational basis of $\mathbbm{C}^d$. It should be noticed here that by the very construction the vectors $\ket{r}_{i,x}$ are mutually orthogonal for any choice of $i$ and $x$.

Now, we are in a position to prove the following fact.
\begin{fakt}
\label{fact:1}
The unitary operators $W_1,W_2, W_{\mathrm{odd}}, W_{\mathrm{ev}}:\mathbbm{C}^d\rightarrow \mathbbm{C}^d$  transform $Z_d,T_{d,m}$ defined in Eq. \eqref{ZdTd} to the ideal measurements given in Eq. \eqref{Obs12} in the following way: $\mathcal{O}_{i,2}=W_i\,Z_d\,W_i^{\dagger}$
and $\mathcal{O}_{i,3}=W_i\,T_{d,m}\,W_i^{\dagger}$
%
%\mathcal{O}_{1,3}=W_1T_{d,m}W_1^{\dagger}$, $\mathcal{O}_{2,2}=W_2Z_dW_2^{\dagger}, %\mathcal{O}_{2,3}=W_2T_{d,m}W_2^{\dagger}$ 
%
for the parties $i=1,2$, and
$\mathcal{O}_{n_{\mathrm{odd}},2}=W_{\mathrm{odd}}\,Z_d\,W_{\mathrm{odd}}^{\dagger}, \mathcal{O}_{n_{\mathrm{odd}},3}=W_{\mathrm{odd}}T_{d,m}W_{\mathrm{odd}}^{\dagger}$, and
$\mathcal{O}_{n_{\mathrm{ev}},2}=W_{\mathrm{ev}}Z_dW_{\mathrm{ev}}^{\dagger}, \mathcal{O}_{n_{\mathrm{ev}},3}=W_{\mathrm{ev}}T_{d,m}W_{\mathrm{ev}}^{\dagger}$ for remaining parties, where the subscript odd and ev refer to odd-numbered and even-numbered parties. The unitary operators are given by
\begin{eqnarray}\label{Unitaries}
W_1&=&\frac{1}{\sqrt{d}}\sum_{i,j=0}^{d-1}(-1)^{\delta_{j,0}}\omega^{-\frac{3i}{2m}+ij+\frac{j}{2}}\ket{i}\!\bra{j},\nonumber\\
W_2&=&\frac{1}{\sqrt{d}}\sum_{i,j=0}^{d-1}(-1)^{\delta_{j,0}}\omega^{-\frac{2i}{m}+ij+\frac{j}{2}}\ket{d-1-i}\!\bra{j}, \nonumber\\
W_{odd}&=&\frac{1}{\sqrt{d}}\sum_{i,j=0}^{d-1}(-1)^{\delta_{j,0}}\omega^{-\frac{i}{m}+ij+\frac{j}{2}}\ket{i}\!\bra{j},\nonumber\\
W_{even}&=&\frac{1}{\sqrt{d}}\sum_{i,j=0}^{d-1}(-1)^{\delta_{j,0}}\omega^{-\frac{i}{m}+ij+\frac{j}{2}}\ket{d-1-i}\!\bra{j}. 
\end{eqnarray}
\end{fakt}

%
%We prove the above statement for $k=1$. For this, we first show that $\{Z_d,T_{d,m}\}$ are unitarily equivalent to the CGLMP measuements.
%
\begin{proof}
Let us consider the spectral decompositions of $Z_d$ and $T_{d,m}$,
\begin{equation}
    Z_d = \sum^{d-1}_{q=0}\w^q \ket{q}\!\bra{q},\qquad T_{d,m}= \sum^{d-1}_{r=0} \w^r \ket{r}\!\bra{r}_{T}. 
\end{equation}

We know that the following spectral decomposition holds, where
$\ket{q}$ form the computational basis in $\mathbbm{C}^d$, whereas $\ket{r}_{Tć}$
are the eigenvectors of $T_{d,m}$ given by
\begin{eqnarray} \label{rTd}
  \ket{r}_{T}= \frac{2\mathbbm{i}}{d}\sin\left(\frac{\pi}{m}\right)\w^{-\frac{d}{2m}}\sum_{q=0}^{d-1}(-1)^{\delta_{q,0}}\frac{\omega^{-\frac{q}{2}}}{1-\omega^{r-q-\frac{1}{m}}}\ket{q}.
  \end{eqnarray}
For clarity, we verify the eigendecomposition of $T_{d,m}$,
\begin{eqnarray}\label{B20}
T_{d,m}\ket{r}_{T}
%&=&\left(\sum_{i=0}^{d-1}\omega^{i+\frac{1}{m}}\proj{i}-\frac{2\mathbbm{i}\sin(\pi/m)}{d}\sum_{i,j=0}^{d-1}(-1)^{\delta_{i,0}+\delta_{j,0}}\omega^{\frac{i+j}{2}+\frac{2-d}{2m}}|i\rangle\!\langle j|\right)\left(\frac{2\mathbbm{i}\sin(\pi/m)\w^{-\frac{d}{2m}}}{d}\sum_{q=0}^{d-1}(-1)^{\delta_{q,0}}\frac{\omega^{-\frac{q}{2}}}{1-\omega^{\left(r-q-\frac{1}{m}\right)}}\ket{q}\right)\nonumber\\
%&=& \frac{2}{d}\left(\sum_{q=0}^{d-1}(-1)^{\delta_{q,0}}\frac{\omega^{\frac{q+1}{2}}}{1-\omega^{r-q-\frac{1}{2}}}\ket{q}-\frac{2}{d}\sum_{q,k=0}^{d-1}(-1)^{\delta_{q,0}}\frac{\omega^{\frac{q+1}{2}}}{1-\omega^{r-k-\frac{1}{2}}}\ket{q}\right)\nonumber\\
=\frac{2\mathbbm{i}}{d}\sin\left(\frac{\pi}{m}\right)\w^{-\frac{d}{2m}}\sum_{q=0}^{d-1}(-1)^{\delta_{q,0}}\omega^{\frac{q}{2}+\frac{1}{m}}\left[\frac{1}{1-\omega^{r-q-\frac{1}{2}}}-\frac{2\mathbbm{i}}{d}\sin\left(\frac{\pi}{m}\right)\w^{-\frac{d}{2m}}\sum_{k=0}^{d-1}\frac{1}{1-\omega^{r-k-\frac{1}{m}}}\right]\ket{q}.\nonumber\\
\end{eqnarray}
Using the formula for the sum of a geometric sequence we have the following relation
\be \label{sumimpW}
\sum_{l=0}^{d-1}\omega^{(r-k-\frac{1}{m})l}=\frac{1-\w^{-\frac{d}{m}}}{1-\omega^{r-k-\frac{1}{m}}}=2\mathbbm{i}\sin\left(\frac{\pi}{m}\right)\frac{\w^{-\frac{d}{2m}}}{1-\omega^{r-k-\frac{1}{m}}},
\ee 
which can later be used to write 
\begin{eqnarray}
\sum_{k=0}^{d-1}\frac{1}{1-\omega^{r-k-\frac{1}{m}}}=\frac{\w^{d/2m}}{2\mathbbm{i}\sin(\pi/m)}\sum_{l=0}^{d-1}\left(\sum_{k=0}^{d-1}\omega^{(r-k-\frac{1}{m})l}\right).
%=\frac{1}{2}\sum_{l=0}^{d-1}\left(\sum_{k=0}^{d-1}\omega^{(r-\frac{1}{2})l}\omega^{-kl}\right).
\end{eqnarray}
Noting that the sum over $k$ is nonzero iff $l=0$, we obtain
\begin{eqnarray} \label{sum10}
\sum_{k=0}^{d-1}\frac{1}{1-\omega^{r-k-\frac{1}{2}}}
%=\frac{1}{2}\sum_{k=0}^{d-1}1
=\frac{d\w^{d/2m}}{2\mathbbm{i}\sin(\pi/m)}.
\end{eqnarray}
Substituting the above relation \eqref{sum10} into Eq. $\eqref{B20}$, we finally have 
\begin{eqnarray}
T_{d,m}\ket{r}_{T} &=& \frac{2\mathbbm{i}}{d}\sin\left(\frac{\pi}{m}\right)\w^{-d/2m}\sum_{q=0}^{d-1}(-1)^{\delta_{q,0}}\omega^{\frac{q}{2}+\frac{1}{m}}\left(\frac{1}{1-\omega^{r-q-\frac{1}{m}}}-1\right)\ket{q} \nonumber  \\
%=\frac{2}{d}\sum_{q=0}^{d-1}(-1)^{\delta_{q,0}}\frac{\omega^{r-\frac{q}{2}}}{1-\omega^{r-q-\frac{1}{2}}}
&=&\omega^r\ket{r}_{T}.
\end{eqnarray}
Thus the vectors $\ket{r}_{T}$ are the eigenvectors of $T_{d,m}$.

Let us now show that the unitary operations \eqref{Unitaries} transform $Z_d$ and $T_{d,m}$ to the optimal measurements $\mathcal{O}_{i,2}$ and $\mathcal{O}_{i,3}$ for any $i=1,\ldots,N$.
To this aim, it is sufficient to show that they transform the eigenvectors of one observable to the eigenvectors of another observable up to a complex number.
Let us first consider $W_1$. The action of its Hermitian conjugation on 
the eigenvectors of $\mathcal{O}_{1,2}$, $\ket{r}_{1,2}$, given explicitly in Eq. (\ref{idealeigen}) can be expressed as
%
% \begin{eqnarray}
%W_1=\frac{1}{\sqrt{d}}\sum_{i,j=0}^{d-1}(-1)^{\delta_{j,0}}\omega^{-\frac{3i}{2m}+ij+\frac{j}{2}}\ke%t{i}\bra{j} ,
% \end{eqnarray}
 %
 %
 %and further
 %
 \begin{eqnarray}
 W_1^{\dagger}\ket{r}_{1,2}
 &=& \frac{1}{d}\sum_{j,q=0}^{d-1}(-1)^{\delta_{j,0}}\omega^{(r-j)q}\omega^{-\frac{j}{2}}\ket{j}.
 \end{eqnarray}
Using the fact that 
\begin{equation}\label{wlasnosc}
    \sum_{q=0}^{d-1}\w^{(r-j)q}=d\delta_{r,j},
\end{equation}
the above simplifies to
\begin{eqnarray}\label{A1toZd}
W_1^{\dagger}\ket{r}_{1,2}
  %=(-1)^{1-\delta_{r,0}}\omega^{-\frac{r}{2}}\ket{q} 
= \w^{\delta_{r,0}-\frac{r}{2}}\ket{r}.
\end{eqnarray}
Since $\ket{r}$ are the eigenvectors of $Z_d$ we thus obtain that $W_1^{\dagger}\mathcal{O}_{1,2}\, W_1=Z_d$. 

Let us now determine the action of $W_1^{\dagger}$ on the eigenvectors of  $\mathcal{O}_{1,3}$.
Using Eqs. (\ref{Unitaries}) and (\ref{idealeigen}) one obtains
 \begin{eqnarray}
W_1^{\dagger}\ket{r}_{1,3}
 &=& \frac{1}{d}\sum_{j,q=0}^{d-1}(-1)^{\delta_{j,0}}\omega^{(r-j-\frac{1}{m})q}\omega^{-\frac{j}{2}}\ket{j} .
 \end{eqnarray}
Taking into account Eqs. \eqref{sumimpW} and (\ref{rTd}) we then have
 \begin{eqnarray} \label{A2toTd}
  W_1^{\dagger}\ket{r}_{1,3}
 &=&\frac{2\mathbbm{i}}{d}\sin\left(\frac{\pi}{m}\right)\w^{-\frac{d}{2m}}\sum_{j=0}^{d-1}(-1)^{\delta_{j,0}}\frac{\omega^{-\frac{j}{2}}}{1-\omega^{r-j-\frac{1}{m}}}\ket{j}\nonumber\\
  &=&\ket{r}_{T_{d,m}}.
 \end{eqnarray}

Let us then consider $W_2$ given by the second formula in (\ref{Unitaries}) and apply $W_2^{\dagger}$ to the eigenvectors of $\mathcal{O}_{2,2}$. This leads us to
%
% \begin{eqnarray}
%W_2=\frac{1}{\sqrt{d}}\sum_{i,j=0}^{d-1}(-1)^{\delta_{j,0}}\omega^{-\frac{2i}{m}+ij+\frac{j}{2}}\ket%{d-1-i}\bra{j},
% \end{eqnarray}
 %
 %
%and further
%
 \begin{eqnarray}
 W_2^{\dagger}\ket{r}_{2,2}
 &=& \frac{1}{d}\sum_{j,q=0}^{d-1}(-1)^{\delta_{j,0}}\omega^{(r-j)q+(d-1)\left(\frac{2}{m}-r\right)-\frac{j}{2}}\ket{j},
 \end{eqnarray}
and, after employing Eq. (\ref{wlasnosc}), to 
 \begin{eqnarray}\label{B1toZd}
  W_2^{\dagger}\ket{r}_{2,2}
  %=(-1)^{1-\delta_{r,0}}\omega^{\left((d-1)(1/2-r)-\frac{r}{2}\right)}\ket{q}
  = (-1)^{\delta_{r,0}}\w^{(d-1)\left(\frac{2}{m}-r\right)-\frac{j}{2}} \ket{r}.
 \end{eqnarray}
Thus, up to some phases, $W_2^{\dagger}$ maps the eigenvectors of $\mathcal{O}_{2,2}$
to those of $Z_d$; in other words, $W_2^{\dagger}\, \mathcal{O}_{2,2}\, W_2=Z_d$.
Analogously, we can write 
 \begin{eqnarray}
 W_2^{\dagger}\ket{r}_{2,3}
 %&=& \left(\frac{1}{\sqrt{d}}\sum_{i,j=0}^{d-1}(-1)^{1-\delta_{j,0}}\omega^{\left(\frac{i}{2}-ij-\frac{j}{2}\right)}\ket{j}\bra{d-1-i}\right)\left(\frac{1}{\sqrt{d}}\sum_{q=0}^{d-1} \omega^{-(r-1)q}\ket{q}\right)\nonumber\\
 &=& \frac{1}{d}\sum_{j,q=0}^{d-1}(-1)^{\delta_{j,0}}\omega^{(r-j-\frac{1}{m})q + (d-1)\left(\frac{2}{m}-r\right)-\frac{j}{2}}\ket{j} ,
 \end{eqnarray}
which after carrying out the sum over $q$ using Eq. \eqref{sumimpW} simplifies to
 \begin{eqnarray}\label{B2toTd}
  W_2^{\dagger}\ket{r}_{2,3}
 % =\frac{2}{d}\sum_{j=0}^{d-1}(-1)^{1-\delta_{j,0}}\omega^{\frac{d-1}{2}}\frac{\omega^{\frac{j}{2}}}{1-\omega^{j+\frac{1}{2}-r}}\ket{j}
  =\omega^{(d-1)\left(\frac{2}{m}-r\right)}\ket{r}_{T}.
 \end{eqnarray}

Next, we look at $W_{\mathrm{odd}}$ defined through the third formula in Eq. (\ref{Unitaries}).
The action of $W_2^{\dagger}$ on the eigenvectors of the second observable $\mathcal{O}_{\mathrm{odd},2}$ of each odd party is given by
%
% \begin{eqnarray}
% W_{odd}&=&\frac{1}{\sqrt{d}}\sum_{i,j=0}^{d-1}(-1)^{\delta_{j,0}}\omega^{-\frac{i}{m}+ij+\frac{j}{2%}}\ket{i}\!\bra{j}
% \end{eqnarray}
%
 \begin{eqnarray}
 W_{\mathrm{odd}}^{\dagger}\ket{r}_{n_{\mathrm{odd}},2}
 &=& \frac{1}{d}\sum_{j,q=0}^{d-1}(-1)^{\delta_{j,0}}\omega^{(r-j)q}\omega^{-\frac{j}{2}}\ket{j},
 \end{eqnarray}
which by using Eq. (\ref{wlasnosc}) can be rewritten as
 \begin{eqnarray}\label{AoddtoZd}
  W_{\mathrm{odd}}^{\dagger}\ket{r}_{\mathrm{odd},2}
  = (-1)^{\delta_{r,0}}\w^{-r/2}\ket{r} .
 \end{eqnarray}
Similarly, we have for $\mathcal{O}_{\mathrm{odd},3}$,
 \begin{eqnarray}
W_{\mathrm{odd}}^{\dagger}\ket{r}_{\mathrm{odd},3}
 &=& \frac{1}{d}\sum_{j,q=0}^{d-1}(-1)^{\delta_{j,0}}\omega^{(r-j-\frac{1}{m})q}\omega^{-\frac{j}{2}}\ket{j}.
 \end{eqnarray}
 which by virtue of Eq. (\ref{sumimpW}) simplifies to 
 \begin{eqnarray} \label{AoddtoTd}
  W_{\mathrm{odd}}^{\dagger}\ket{r}_{\mathrm{odd},3}
 %=\frac{2\mathbbm{i}\sin(\pi/m)\w^{-\frac{d}{2m}}}{d}\sum_{j=0}^{d-1}(-1)^{\delta_{j,0}}\frac{\omega^{-\frac{j}{2}}}{1-\omega^{r-j-\frac{1}{m}}}\ket{j}
  =\ket{r}_{T}.
 \end{eqnarray}

Let us finally consider $W_{\mathrm{ev}}$ given by the forth equation of (\ref{Unitaries}).
We have
%
% \begin{eqnarray}
%W_{even}=\frac{1}{\sqrt{d}}\sum_{i,j=0}^{d-1}(-1)^{\delta_{j,0}}\omega^{-\frac{i}{m}+ij+\frac{j}{2}}%\ket{d-1-i}\bra{j},
% \end{eqnarray}
 %
 %
 \begin{eqnarray}
 W_{\mathrm{ev}}^{\dagger}\ket{r}_{\mathrm{ev},2}
 &=& \frac{1}{d}\sum_{j,q=0}^{d-1}(-1)^{\delta_{j,0}}\omega^{(r-j)q+(d-1)\left(\frac{2}{m}-r\right)-\frac{j}{2}}\ket{j},
 \end{eqnarray}
which after performing the summation over $q$ simplifies to
 \begin{eqnarray}\label{BeventoZd}
  W_{\mathrm{ev}}^{\dagger}\ket{r}_{\mathrm{ev},2}
  %=(-1)^{1-\delta_{r,0}}\omega^{\left((d-1)(1/2-r)-\frac{r}{2}\right)}\ket{q}
  =(-1)^{\delta_{r,0}} \w^{(d-1)\left(\frac{2}{m}-r\right)-\frac{r}{2}} \ket{r}.
 \end{eqnarray}
Consequently, $W_{\mathrm{ev}}^{\dagger}\,\mathcal{O}_{\mathrm{ev},2}\,W_{\mathrm{ev}}=Z_d$.

Then, for the third observable $\mathcal{O}_{\mathrm{ev},3}$ we have
 \begin{eqnarray}
 W_{\mathrm{ev}}^{\dagger}\ket{r}_{\mathrm{ev},3}
 &=& \frac{1}{d}\sum_{j,q=0}^{d-1}(-1)^{\delta_{j,0}}\omega^{(r-j-\frac{1}{m})q + (d-1)\left(\frac{2}{m}-r\right)-\frac{j}{2}}\ket{j},
 \end{eqnarray}
which by using Eq. \eqref{sumimpW} reduces to
 \begin{eqnarray}\label{BeventoTd}
  W_{\mathrm{ev}}^{\dagger}\ket{r}_{\mathrm{ev},3}
 % =\frac{2}{d}\sum_{j=0}^{d-1}(-1)^{1-\delta_{j,0}}\omega^{\frac{d-1}{2}}\frac{\omega^{\frac{j}{2}}}{1-\omega^{j+\frac{1}{2}-r}}\ket{j}
  =\omega^{(d-1)\left(\frac{2}{m}-r\right)}\ket{r}_{T},
 \end{eqnarray}
implying that $W_{\mathrm{ev}}^{\dagger}\,\mathcal{O}_{\mathrm{ev},3}\,W_{\mathrm{ev}}=T_{d,m}$.
This completes the proof.
%
%Altogether, Eqs. \eqref{A1toZd}, \eqref{A2toTd}, \eqref{B1toZd}, \eqref{B2toTd}, \eqref{AoddtoZd}, %\eqref{AoddtoTd}, \eqref{BeventoZd} and \eqref{BeventoTd} complete the proof.
\end{proof}

Now, we can show that the measurements $A_{i,\alpha}$ for all $i,\alpha$ are equivalent to the optimal measurements \eqref{Obs12} and \eqref{Obsoddeven}. For this, we recall the relations obtained in \eqref{SOSrelb} and \eqref{SOSrelb1} for $k=1$,
\begin{eqnarray}\label{SOS28}
R_{n,\alpha}^{(1)}=\mu_{\alpha,1}^*A_{n,2}+\nu_{\alpha,1}^*A_{n,\alpha+2}+\tau_{\alpha,1}A_{n,\alpha+3}=0
\end{eqnarray}
when $n$ is odd, and when $n$ is even
\begin{eqnarray}\label{SOS29}
R_{n,\alpha}^{(1)}=\mu_{\alpha,k}A_{n,2}^{-1}+\nu_{\alpha,1}A_{n,\alpha+2}^{-1}+\tau_{\alpha,1}A_{n,\alpha+3}^{-1}=0
\end{eqnarray}
where the coefficients $\mu_{\alpha,1}, \nu_{\alpha,1}$ and $\tau_{\alpha,1}$ are given in Eq. \eqref{Rcoef1} and Eq. \eqref{Rcoef2}. A key observation here is that 
the ideal observables $\mathcal{O}_{n,\alpha}$ are known to maximally violate the above Bell inequality and thus satisfy the relations (\ref{SOS28}) and (\ref{SOS29}). Next, we choose $\alpha=1$ and we observe from \eqref{SOS28} and \eqref{SOS29} that for all $n$,
\begin{eqnarray}
A_{n,4}=-\frac{1}{\tau_{1,1}}\left(\mu_{1,1}^*A_{n,2}+\nu_{1,1}^*A_{n,3}\right),
\end{eqnarray}
where we used the fact that $A_{i,\alpha}^{-1}=A_{i,\alpha}^{\dagger}$. We showed in Fact \ref{fact:1} that $A_{n,2}$ and $A_{n,3}$ are equivalent to the optimal measurements $\mathcal{O}_{n,2}\otimes\I_{n}'$ and $\mathcal{O}_{n,3}\otimes\I_{n}'$ given in Eqs. \eqref{Obs12} and \eqref{Obsoddeven}. As a consequence, $A_{n,4}$ is equivalent to $\mathcal{O}_{n,4}\otimes\I_{n}'$ up to a unitary transformation. Similarly, we can put $\alpha=2$ in \eqref{SOS28} and \eqref{SOS29} and conclude that for all $n$,
\begin{eqnarray}
A_{n,5}=-\frac{1}{\tau_{2,1}}\left(\mu_{2,1}^*A_{n,2}+\nu_{2,1}^*A_{n,4}\right).
\end{eqnarray}
This implies that $A_{n,5}$ is equivalent to $\mathcal{O}_{n,5}\otimes\I_{n}'$ up to some unitary transformation. We continue in a similar manner, and conclude that there exist local unitary transformations
%
%\begin{equation}
 $   U_{i}: \mathcal{H}_{i} \rightarrow \mathbbm{C}^d \otimes \mathcal{H}_{i}$
such that 
\begin{eqnarray}\label{AiBimain3}
U_{n}\, A_{n,\alpha}\, U^{\dagger}_{n} &=& \mathcal{O}_{n,\alpha} \otimes \I_{n} \quad \forall n,\alpha.
\end{eqnarray}  
For a note, we explicitly calculate $A_{1,\alpha+3}$ in \eqref{SOSrelb} for $k=1$ given that $A_{1,2}=\mathcal{O}_{1,2}\otimes\I_{1}'$ and $A_{1,\alpha+2}=\mathcal{O}_{1,\alpha+2}\otimes\I_{1}'$,
\begin{eqnarray}
U_1A_{1,\alpha+3}U_1^{\dagger}=-\frac{\mu_{\alpha,1}^*}{\tau_{\alpha,1}}\mathcal{O}_{1,2}\otimes\I_{1}'-\frac{\nu_{\alpha,1}^*}{\tau_{\alpha,1}}\mathcal{O}_{1,\alpha+2}\otimes\I_{1}'. 
\end{eqnarray}
Plugging in the values of $\mu_{\alpha,1},\nu_{\alpha,1}, \tau_{\alpha,1}$ from \eqref{Rcoef1} for $\alpha=1,2,\ldots,m-3$, and simplifying
\begin{eqnarray}
U_1A_{1,\alpha+3}U_1^{\dagger}=\sum_{i=0}^{d-1}\w^{\gamma_m(2)(1-d\delta_{i,d-1})}\w^{\frac{2-d}{2m}}\w^{\frac{\alpha(1-d\delta_{i,d-1})}{m}}\left(-\sin(\pi/m)\w^{-\frac{\alpha d(1-2\delta_{i,d-1})}{2m}}+\sin(\pi(\alpha+1)/m)\right)\frac{1}{\sin(\pi\alpha/m)}\nonumber\\\ket{i}\bra{i+1}\otimes\I_{\mathcal{A}''_1}\nonumber\\
\end{eqnarray}
which on further simplification gives us
\begin{eqnarray}
U_1A_{1,\alpha+3}U_1^{\dagger}=\left(\sum_{i=0}^{d-1}\w^{\gamma_m(3)(1-d\delta_{i,d-1})}\w^{\frac{\alpha(1-d\delta_{i,d-1})}{m}}\ket{i}\bra{i+1}\right)\otimes\I_{\mathcal{A}''_1}=\mathcal{O}_{1,\alpha+3}\otimes\I_{\mathcal{A}''_1}.
\end{eqnarray}
Now plugging in the values of $\mu_{m-2,1},\nu_{m-2,1}, \tau_{m-2,1}$ from \eqref{Rcoef2}, and simplifying
\begin{eqnarray}
U_1A_{1,m+1}U_1^{\dagger}=\left(\sum_{i=0}^{d-1}\w^{-\frac{(1-d\delta_{i,d-1})}{2m}+1}\left(\w^{\frac{d-2}{2m}}\w^{\frac{2(1-d\delta_{i,d-1})}{m}}+\w^{\frac{2-d}{2m}}\right)\frac{1}{2\cos(\pi/m)}\ket{i}\bra{i+1}\right)\otimes\I_{\mathcal{A}''_1}
\end{eqnarray}
which on further simplification gives,
\begin{eqnarray}
U_1A_{1,m+1}U_1^{\dagger}=\w A_{1,1}=\w\left(\w^{\gamma_m(1)(1-d\delta_{i,d-1})}\w^{\frac{\alpha(1-d\delta_{i,d-1})}{m}}\ket{i}\bra{i+1}\right)\otimes\I_{\mathcal{A}''_1}=\w\mathcal{O}_{1,1}\otimes\I_{\mathcal{A}''_1}.
\end{eqnarray}
For all the other $n$, $A_{n,\alpha_n}$ can be computed in the similar way. This completes the first part of the proof which involves finding the observables and finally, we can proceed towards the last step of the proof which involves finding the state which maximally violates the Bell inequality $\langle\mathcal{I}_{N,m,d}\rangle$ \eqref{eqASTAop}.

\textbf{The state.} Let us first look at the quantity $\overline{A}_{1,\alpha_1}^{(1)}$ from \eqref{SOSrela}. As derived in \eqref{AiBimain}, we expand $\overline{A}_{1,\alpha_1}^{(1)}$ by putting in the optimal measurements from \eqref{Obs12}.
\begin{eqnarray}
U_1\overline{A}_{1,\alpha}^{(1)}U_1^{\dagger}&=&a_1O_{1,\alpha}+a_1^*O_{1,\alpha+1}\nonumber\\&=&\left[\sum_{i=0}^{d-2}\omega^{\gamma_m(\alpha)}\left(a_1+a_1^*\omega^{\frac{1}{m}}\right)\ket{i}\!\bra{i+1}+\omega^{(1-d)\gamma_2(\alpha)}\left(a_1+a_1^*\omega^{-\frac{d-1}{m}}\right)\ket{d-1}\!\bra{0}\right]\otimes\I_{n}'.
\end{eqnarray}
Simplifying and using the fact that $a_1+a_1^*\omega^{1/m}=\w^{1/2m}$ and $a_1+a_1^*\omega^{-(d-1)/m}=\w^{-(d-1)/2m}$, we get
\begin{eqnarray}
U_1\overline{A}_{1,\alpha}^{(1)}U_1^{\dagger}=\left[\sum_{i=0}^{d-2}\omega^{\zeta_m(\alpha)}\ket{i}\!\bra{i+1}+\omega^{-(d-1)\zeta_m(\alpha)}\ket{d-1}\!\bra{0}\right]\otimes\I_{n}'
\end{eqnarray}
where we used the fact that $\gamma_m(x)+1/2m=\zeta_m(x)$ [cf. Eq. \eqref{measupara}]. 
For ease of calculation, we first look at how each of the measurements from \eqref{SOSrel2} act on any vector from $\mathbbm{C}^d\otimes \mathcal{H}_{1}'$ of the form $\ket{j}\ket{\phi}$, where $\ket{j}$ is an element of the computational basis of $\mathbbm{C}^d$ whereas $\ket{\phi}$
is an arbitrary vector from $\mathcal{H}_1'$,
\begin{eqnarray}
U_1\overline{A}_{1,x}^{(1)}U_1^{\dagger}\ket{j}\ket{\phi}&=&\omega^{(1-d\delta_{j,0})(x/m)}\ket{j-1}\ket{\phi},
\nonumber\\
%
%\overline{A}_{1,x}^{(1)}\ket{j}&=&\left(\sum_{i=0}^{d-1}\omega^{(1-d\delta_{i,d-1})\zeta_m(x)}\ket{i%}\delta_{i+1,j}\right)\otimes\I_{1}',\nonumber\\
%%%%%%%%%%%%%%%%%%%%%
U_2A_{2,x}^{-1}U_2^{\dagger}\ket{j}\ket{\phi}&=&\omega^{-(1-d\delta_{j,0})(x/m)}\ket{j-1}\ket{\phi},\nonumber\\
%
%A_{2,x}^{-1}\ket{j}&=&\left(\sum_{i=0}^{d-1}\omega^{-(1-d\delta_{i,d-1})\zeta_m(x)}\ket{i}\delta_{i+%1,j}\right)\otimes\I_{1}',\nonumber\\
%%%%%%%%%%%%%%%%%%
U_{n_{\mathrm{odd}}}A_{n_{\mathrm{odd}},x}U_{n_{odd}}^{\dagger}\ket{j}\ket{\phi}&=&\omega^{(1-d\delta_{j,0})\theta_m(x)}\ket{j-1}\ket{\phi},\nonumber\\
%%%%%%%%%%%%%%%%%
U_{n_{\mathrm{ev}}}A_{n_{\mathrm{ev}},x}^{-1}U_{n_{ev}}^{\dagger}\ket{j}\ket{\phi}&=&\omega^{-(1-d\delta_{j,0})\theta_m(x)}\ket{j-1}\ket{\phi},
\end{eqnarray}
where $\ket{-1}\equiv \ket{d-1}$.

Having determined the action of the measurements on the elements of the standard basis, 
let us then decompose the state $\ket{\tilde{\psi}_N}=U_1\otimes \ldots\otimes U_N\ket{\psi_N}$ as
\begin{eqnarray}\label{genstate}
\ket{\tilde{\psi}_{N}}=\sum_{i_1,\ldots, i_N=0}^{d-1}\ket{i_1,\ldots, i_N}\ket{\psi_{i_1,\ldots, i_N}}
\end{eqnarray}
for some, in general unnormalized, vectors $\ket{\psi_{i_1,\ldots ,i_N}}\in\mathcal{H}_{1}'\otimes\ldots\otimes\mathcal{H}_N'$,
and consider the relations \eqref{SOSrel2} for $\alpha_1=\alpha_2=\ldots=\alpha_N=1$ and $k=1$.
Taking into account that $\theta_m(1)=0$, this relation gives
\begin{eqnarray}\label{stateeq1.0}
\sum_{i_1,\ldots, i_N=0}^{d-1}\w^{\frac{d}{m}\left(\delta_{i_2,0}-\delta_{i_1,0}\right)}\ket{i_1-1}\ldots \ket{i_N-1}\ket{\psi_{i_1,\ldots, i_N}}=\sum_{i_1,\ldots, i_N=0}^{d-1}\ket{i_1,\ldots, i_N}\ket{\psi_{i_1,\ldots, i_N}},
\end{eqnarray}
from which we directly obtain that
\begin{eqnarray}\label{state1.1}
\forall i_1,\ldots,i_N\qquad\w^{\frac{d}{m}\left(\delta_{i_2,0}-\delta_{i_1,0}\right)}\ket{\psi_{i_1,\ldots, i_N}}=\ket{\psi_{i_1-1,\ldots, i_N-1}}.
\end{eqnarray}
Again, considering the relations \eqref{SOSrel2} for $\alpha_1=2$ and $\alpha_2=\ldots=\alpha_N=1$ with $k=1$, we have
\begin{eqnarray}\label{state1.2}
\forall i_1,\ldots,i_N\qquad \w^{\frac{2d}{m}\left(\delta_{i_2,0}-\delta_{i_1,0}\right)}\ket{\psi_{i_1,\ldots, i_N}}=\ket{\psi_{i_1-1,\ldots, i_N-1}}.
\end{eqnarray}
Simultaneously, solving the above equations \eqref{state1.1} and \eqref{state1.2}, we have the following conditions. First, when $\delta_{i_2,0}=\delta_{i_1,0}$,
\begin{eqnarray}\label{state1.3}
\ket{\psi_{i_1,i_2\ldots, i_N}}=\ket{\psi_{i_1-1,i_2-1,\ldots, i_N-1}} \quad \text{for}\quad i_1,i_2=1,2,\ldots,d-1\ \ \text{or}\ \  i_1=i_2=0.
\end{eqnarray}
and for all $i_3,i_4,\ldots,i_N$. Second, when $\delta_{i_1,0}\ne\delta_{i_2,0}$,
\begin{eqnarray}\label{state1.4}
\ket{\psi_{i_1,0,i_3,\ldots, i_N}}=0, \quad \ket{\psi_{0,i_2,\ldots, i_N}}=0 \quad \text{for}\quad i_1,i_2=1,2,\ldots,d-1
\end{eqnarray}
and for all $i_3,i_4,\ldots,i_N$. Now consider \eqref{state1.3} for $i_2=1$ and $i_1\ne 1$,
\begin{eqnarray}
\ket{\psi_{i_1,1\ldots, i_N}}=\ket{\psi_{i_1-1,0,\ldots, i_N-1}}=0% \quad \text{for}\quad i_1\ne1.
\end{eqnarray}
Again considering \eqref{state1.3} for $i_2=2$ and $i_1\ne 2$,
\begin{eqnarray}
\ket{\psi_{i_2,2,\ldots, i_N}}=\ket{\psi_{i_1-1,1,\ldots, i_N-1}}=0% \quad \text{for}\quad i_1\ne2.
\end{eqnarray}
Continuing in a similar way, we have that
\begin{eqnarray}\label{state1.51}
\ket{\psi_{i_1,i_2,\ldots, i_N}}=0\quad \quad \forall i_1,i_2,\ldots,i_N\ \ \text{s.t.}\ \ i_1\ne i_2
\end{eqnarray}
and,
\begin{eqnarray}\label{state1.5}
\ket{\psi_{i_2-1,i_2-1,i_3-1,\ldots, i_N-1}}=\ket{\psi_{i_2,i_2,i_3,\ldots, i_N}}\qquad \forall i_2,i_3,\ldots,i_N.
\end{eqnarray}
Using the above conditions \eqref{state1.5}, \eqref{state1.51} and considering the relations \eqref{SOSrel2} for $\alpha_1=\alpha_3=\ldots=\alpha_N=1$ and $\alpha_2=2$, we arrive at the following condition
\begin{eqnarray}\label{eq132}
 \forall i_2,\ldots,i_N\qquad\w^{\frac{d}{m}\left(\delta_{i_2,0}-\delta_{i_3,0}\right)}\ket{\psi_{i_2,i_2,i_3\ldots, i_N}}=\ket{\psi_{i_2-1,i_2-1,i_3-1,\ldots, i_N-1}}.
\end{eqnarray}
For the case, when $\delta_{i_2,0}=\delta_{i_3,0}$ we have
\begin{eqnarray}\label{state1.6}
\ket{\psi_{i_2,i_2,i_3,\ldots, i_N}}=\ket{\psi_{i_2-1,i_2-1,i_3-1,\ldots, i_N-1}} \quad \text{for}\quad i_2,i_3=1,2,\ldots,d-1\ \ \text{or}\ \  i_2=i_3=0.
\end{eqnarray}
and for all $i_4,i_5\ldots,i_N$. Second when $\delta_{i_2,0}\ne\delta_{i_3,0}$, using \eqref{state1.3} and \eqref{eq132} we can conclude that
\begin{eqnarray}\label{state133}
\ket{\psi_{0,0,i_3,\ldots, i_N}}=0, \quad \ket{\psi_{i_2,i_2,0,\ldots, i_N}}=0 \quad \text{for}\quad i_2,i_3=1,2,\ldots,d-1
\end{eqnarray}
and for all $i_4,i_5\ldots,i_N$. Again considering \eqref{state133} for $i_2=1$ and $i_3\ne 1$,
\begin{eqnarray}
\ket{\psi_{1,1,i_3,\ldots, i_N}}=\ket{\psi_{0,0,i_3,\ldots, i_N-1}}=0% \quad \text{for}\quad i_3\ne1.
\end{eqnarray}
Again considering \eqref{state133} for $i_2=2$ and $i_3\ne 2$,
\begin{eqnarray}
\ket{\psi_{2,2,i_3,\ldots, i_N}}=\ket{\psi_{1,1,i_3,\ldots, i_N-1}}=0% \quad \text{for}\quad i_3\ne2.
\end{eqnarray}
Continuing in a similar way, we have that
\begin{eqnarray}\label{state1.7}
\ket{\psi_{i_2,i_2,i_3,\ldots, i_N}}=0\qquad \forall i_2,i_3,\ldots,i_N\ \ \text{s.t.}\ \ i_2\ne i_3
\end{eqnarray}
and,
\begin{eqnarray}\label{state1.71}
\ket{\psi_{i_2-1,i_2-1,i_2-1,\ldots, i_N-1}}=\ket{\psi_{i_2,i_2,i_2,\ldots ,i_N}}\qquad \forall i_2,i_4,\ldots,i_N.
\end{eqnarray}
Using the above conditions \eqref{state1.7} and \eqref{state1.71}, we proceed in a similar manner by again considering the relations \eqref{SOSrel2} for $\alpha_1=\alpha_2=\alpha_4=\ldots=\alpha_N=1$ and $\alpha_{3}=2$ and arrive at
\begin{eqnarray}\label{eq139}
\w^{\frac{d}{m}\left(\delta_{i_4,0}-\delta_{i_2,0}\right)}\ket{\psi_{i_2,i_2,i_2,i_4,\ldots, i_N}}=\ket{\psi_{i_2-1,i_2-1,i_2-1,i_4-1,\ldots, i_N-1}}\quad \forall i_2,i_4,\ldots,i_N.
\end{eqnarray}
For the case, when $\delta_{i_2,0}=\delta_{i_4,0}$ we have
\begin{eqnarray}\label{state1.8}
\ket{\psi_{i_2,i_2,i_2,i_4,\ldots, i_N}}=\ket{\psi_{i_2-1,i_2-1,i_2-1,i_4,\ldots, i_N-1}} \quad \text{for}\ \ i_2,i_4=1,2,\ldots,d-1\ \ \text{or}\ \  i_2=i_4=0.
\end{eqnarray}
and for all $i_5,i_6,\ldots,i_N$. For the case when $\delta_{i_2,0}\ne\delta_{i_4,0}$  along with \eqref{state1.3} and \eqref{eq139}, we have
\begin{eqnarray}\label{state141}
\ket{\psi_{0,0,0,i_4,\ldots, i_N}}=0, \quad \ket{\psi_{i_2,i_2,i_2,0,\ldots, i_N}}=0 \quad \text{for}\quad i_2,i_4=1,2,\ldots,d-1
\end{eqnarray}
and for all $i_5,i_6\ldots,i_N$. In a similar manner as concluded above, we again have that
\begin{eqnarray}
\ket{\psi_{i_2,i_2,i_2,i_4,\ldots, i_N}}=0\quad \quad \forall i_2,i_4\ldots,i_N\ \ \text{s.t.}\ \ i_2\ne i_4
\end{eqnarray}
and,
\begin{eqnarray}
\ket{\psi_{i_2-1,i_2-1,i_2-1,i_2-1\ldots, i_N-1}}=\ket{\psi_{i_2,i_2,i_2,i_2,\ldots, i_N}}\quad \forall i_2,i_5,\ldots,i_N.
\end{eqnarray}
We proceed in a similar manner, considering $N-1$ different equations with $\alpha_n=2$ for all $n\ne N$ with the rest of coefficients as, $\alpha_1=\alpha_2=\alpha_3=\ldots=\alpha_N=1$  and conclude that the only terms among $\ket{\psi_{i_1,i_2,i_3,\ldots, i_N}}$ which are non-zero are related as,
\begin{eqnarray}
\ket{\psi_{i-1,i-1,i-1,\ldots, i-1}}=\ket{\psi_{i,i,i,\ldots, i}}\quad \forall i.
\end{eqnarray}
As a consequence, with the proper normalization we can conclude that
\begin{eqnarray}
U_1\otimes\ldots \otimes U_N\ket{\psi_{N}}=\left(\frac{1}{\sqrt{d}}\sum_{i=0}^{d-1}\ket{i}^{\otimes N}\right)\otimes\ket{\psi_{0,0,\ldots,0}}
\end{eqnarray}
which is the $N-$partite GHZ state of local dimension $d$ along with some uncorrelated auxiliary state, denoted by $\ket{\psi_{0,0,\ldots,0}}$. This finally completes the proof of our self-testing scheme.
\end{proof}

%\begin{thebibliography}{11}

%\bibitem{ACTA} R. Augusiak, A. Salavrakos, J. Tura and A Ac\'in, New J. Phys. \textbf{21} 113001 (2019).

%\bibitem{Sarkar} S. Sarkar, D. Saha, J. Kaniewski, R. Augusiak, npj Quantum Inf. \textbf{7}, 151 (2021). 

%\end{thebibliography} 

\end{document}